%% file: ms.tex
\documentclass[twocolumn, tighten, numberedappendix]{aastex62}

\usepackage[normalem]{ulem}

\turnoffedit

\usepackage{lipsum}
\usepackage{amsmath}
\usepackage[utf8x]{inputenc}

\usepackage[encapsulated]{CJK}
\usepackage{ucs}
\newcommand{\cntext}[1]{\begin{CJK}{UTF8}{gbsn}#1\end{CJK}}

\usepackage{newtxtext}
\usepackage[varg]{newtxmath}

\usepackage{bm}

\usepackage{ulem}

\shortauthors{Watts et al.}

\usepackage{devanagari}

\begin{document}

\input{commands.tex}

\title{A Projected Estimate of the Reionization Optical Depth
using the CLASS Experiment's\\  Sample Variance Limited  E-Mode Measurement}
\shorttitle{CLASS $\tau$ projection}

\author[0000-0002-5437-6121]{Duncan J.~Watts}
\affiliation{Department of Physics and Astronomy, Johns Hopkins University, 3701 San Martin Drive, Baltimore, MD 21218, USA}

\author[0000-0001-9269-5046]{Bingjie Wang (\cntext{王冰洁}\!)}
\affiliation{Department of Physics and Astronomy, Johns Hopkins University,
3701 San Martin Drive, Baltimore, MD 21218, USA}


\author{Aamir Ali}
\affiliation{Department of Physics,
University Of California,
Berkeley, CA 94720, USA}
\affiliation{Department of Physics and Astronomy, Johns Hopkins University,
3701 San Martin Drive, Baltimore, MD 21218, USA}

\author[0000-0002-8412-630X]{John W.~Appel}
\affiliation{Department of Physics and Astronomy, Johns Hopkins University, 
3701 San Martin Drive, Baltimore, MD 21218, USA}

\author[0000-0001-8839-7206]{Charles L.~Bennett}
\affiliation{Department of Physics and Astronomy, Johns Hopkins University, 
3701 San Martin Drive, Baltimore, MD 21218, USA}

\author[0000-0003-0016-0533]{David T.~Chuss}
\affiliation{Department of Physics, Villanova University, 800 Lancaster Avenue, Villanova, PA 19085, USA
}

\author[0000-0002-1708-5464]{Sumit Dahal ({\dn \7{s}Emt dAhAl})}
\affiliation{Department of Physics and Astronomy, Johns Hopkins University,
3701 San Martin Drive, Baltimore, MD 21218, USA}

\author[0000-0001-6976-180X]{Joseph R.~Eimer}
\affiliation{Department of Physics and Astronomy, Johns Hopkins University,
3701 San Martin Drive, Baltimore, MD 21218, USA}

\author[0000-0002-4782-3851]{Thomas~Essinger-Hileman}
\affiliation{Goddard Space Flight Center, 8800 Greenbelt Road, Greenbelt, MD 20771, USA}

\author{Kathleen Harrington}
\affiliation{Department of Physics and Astronomy, Johns Hopkins University,
3701 San Martin Drive, Baltimore, MD 21218, USA}

\author{Gary Hinshaw}
\affiliation{Department of Physics and Astronomy, University of British Columbia, 6224 Agricultural Road, Vancouver, BC V6T 1Z1, Canada}

\author{Jeffrey Iuliano}
\affiliation{Department of Physics and Astronomy, Johns Hopkins University,
3701 San Martin Drive, Baltimore, MD 21218, USA}

\author{Tobias A.~Marriage}
\affiliation{Department of Physics and Astronomy, Johns Hopkins University, 
3701 San Martin Drive, Baltimore, MD 21218, USA}

\author{Nathan J.~Miller}
\affiliation{Department of Physics and Astronomy, Johns Hopkins University, 
3701 San Martin Drive, Baltimore, MD 21218, USA}
\affiliation{Goddard Space Flight Center, 8800 Greenbelt Road, Greenbelt, MD 20771, USA}

\author{Ivan L.~Padilla}
\affiliation{Department of Physics and Astronomy, Johns Hopkins University,
3701 San Martin Drive, Baltimore, MD 21218, USA}

\author{Lucas Parker}
\affiliation{Space and Remote Sensing, MS D436, Los Alamos National Laboratory,
Los Alamos, NM 87544, USA}
\affiliation{Department of Physics and Astronomy, Johns Hopkins University, 3701 San Martin Drive, Baltimore, MD 21218, USA}

\author[0000-0002-4436-4215]{Matthew Petroff}
\affiliation{Department of Physics and Astronomy, Johns Hopkins University,
3701 San Martin Drive, Baltimore, MD 21218, USA}

\author{Karwan Rostem}
\affiliation{Goddard Space Flight Center, 8800 Greenbelt Road, Greenbelt, MD 20771, USA}

\author[0000-0002-7567-4451]{Edward J.~Wollack}
\affiliation{Goddard Space Flight Center, 8800 Greenbelt Road, Greenbelt, MD 20771, USA}

\author[0000-0001-5112-2567]{Zhilei Xu (\cntext{徐智磊}\!)}
\affiliation{Department of Physics and Astronomy, University of Pennsylvania, 
209 South 33rd Street, Philadelphia, PA 19104, USA}
\affiliation{Department of Physics and Astronomy, Johns Hopkins University, 
3701 San Martin Drive, Baltimore, MD 21218, USA}

\correspondingauthor{Duncan J.~Watts}
\email{dwatts@jhu.edu}

\begin{abstract}
\added{We analyze simulated maps of the Cosmology Large Angular Scale Surveyor (CLASS) experiment and recover a nearly cosmic variance limited estimate of the reionization optical depth $\tau$. We 
    use a  power spectrum-based likelihood to simultaneously clean foregrounds and
    estimate cosmological parameters in multipole space.
    Using software specifically designed to constrain $\tau$, the amplitude of scalar fluctuations $A_s$, and the tensor-to-scalar ratio $r$, we demonstrate that the CLASS experiment will be able to estimate $\tau$ within a factor of two of the cosmic variance limit allowed by full-sky cosmic microwave background polarization measurements. Additionally, we discuss the role of CLASS's $\tau$ constraint in conjunction with gravitational lensing of the CMB on obtaining a
    $\gtrsim4\sigma$ measurement of the sum of the neutrino masses.
    }
\end{abstract}

\keywords{cosmic background radiation -- cosmological parameters -- early
universe -- gravitational waves -- inflation}

\received{January 4, 2018}
\revised{May 22, 2018}
\accepted{July 8, 2018}
\submitjournal{\apj}

\section{Introduction}

Measurements of the cosmic microwave background (CMB) have tightly constrained
the properties of the large-scale observable universe, with the reionization optical depth $\tau$ left as the worst-determined fundamental $\Lambda$CDM parameter \citep{wmapfinal,planck}. The
importance of polarization measurements has become more critical as the \planck\
experiment has measured the unpolarized temperature anisotropy over the full sky
to its sample variance limit up to a resolution of $\theta\gtrsim7'$
($\ell\lesssim1600$) \citep[\S3.8 of][albeit with potential complications, see \citealt{addison}]{plancklike15}. 
At sub-degree angular scales
($\ell\gtrsim200$), 
polarization power is
sourced by primordial scalar fluctuations with extra correlations induced by
gravitational lensing
\citep[e.g.][]{principalpower,Galli2014, sptEmode,actpollens}.
At larger angular scales, gradient-like
E-mode polarization measurements can tightly constrain the reionization optical depth 
$\tau$ via the rough scaling
$C_{2\leqslant\ell\leqslant20}^\mathrm{EE}\propto\tau^2$ \citep{pol_mask}, while we can use the
curl-like B-mode polarization measurements to constrain the amplitude of
stochastic gravitational waves that the inflationary paradigm predicts, whose
amplitude is parameterized by the ratio $r$ of tensor-to-scalar fluctuations in
the metric \citep{kazanas,starobinsky, einhorn,  guth, mukhanov,albrecht, linde,  kks,seljzald}.
The CLASS experiment is uniquely and specially designed to constrain $r$ and $\tau$ by recovering the largest scale fluctuations of the polarized CMB across 70\% of the sky \citep{class, class_spie, 2016SPIE.9914E..1KH}.

The reionization optical depth $\tau$ is the total free electron opacity to the surface of last scattering, 
\begin{equation}
    \tau=\int_{t_\mathrm{lss}}^{t_0}n_\mathrm e(t)\sigma_\mathrm T c\ud t,
\end{equation}
where $n_\mathrm e(t)$ is the average number density of free electrons from the time of last scattering $t_\mathrm{lss}$ to today $t_0$ and $\sigma_\mathrm T$ is the Thomson scattering cross section. For $\tau \ll 1$, the reionization optical depth is the probability that a CMB photon was scattered by free electrons from reionization. The redshift of reionization can be defined if one assumes that $n_\mathrm e(t)$ is nearly a step function, but
it is likely that reionization was an extended process, with evidence of significant contributions to $\tau$ up to $z\sim16$
\citep{2017PhRvD..95b3513H}.

From measurements of QSO
absorption lines via the Gunn--Peterson effect \citep{gunnpeterson}, we know that the universe was ionized by redshift $z=6$,
corresponding to a lower limit of $\tau\gtrsim0.038$ if we assume instantaneous
reionization \citep{fan, plancktau}. The  quantity
$\tau$ can be constrained using measurements
of the temperature-E-mode cross-correlation and the E-mode auto-correlation
$C_\ell^\mathrm{TE}$ and $C_\ell^\mathrm{EE}$ at the largest angular scales.
The \planck\ and \wmap\ measurements are limited in precision by sample variance
in the $C_\ell^\mathrm{TE}$ case, and by instrumental noise  and systematic
effects in the
$C_\ell^\mathrm{EE}$ case, with the latest limits from \planck\ giving
$\tau\sim0.06\pm0.01$ \citep{planck_redsys, plancktau}, although the amplitude of unexplained large-scale
signals in the \planck\ maps create extra uncertainty and potential biases
in this measurement \citep{weiland}.
It is possible to obtain this constraint using only temperature anisotropy, CMB lensing, and baryon acoustic oscillation (BAO) data as an independent check. In particular, \texttt{PlanckTT+lensing+BAO} data constrain $\tau=0.067\pm0.016$ \citep[\S3.4]{planck}  and \texttt{WMAPTT+lensing+BAO} data imply $\tau=0.066\pm0.02$ \citep[\S5]{weiland}. These constraints are independent of CMB polarization data.

Free streaming of massive neutrinos reduces the amplitude of matter fluctuations at small scales.
For testing extensions to \lcdm, a measurement of $\tau$
is necessary to reduce degeneracies between the clustering amplitude at $8\,h^{-1}\mpc$, the physical cold dark matter density, and the sum of the neutrino masses  ($\sigma_8$, $\Omega_ch^2$, and $\sum
m_\nu$, respectively) \citep{allison, Liu2015}.
The measurement of neutrino masses is especially tantalizing since
current upper limits are only a few standard deviations away from the lower limit implied by
solar neutrino oscillation measurements \citep{s4book}.

The relevant polarized foregrounds, thermal dust and synchrotron emission,
dominate at large scales with their angular power spectra approximated by power laws
${C_\ell^\mathrm{dust}\propto\ell^{-2.53}}$ and
$C_\ell^\mathrm{sync}\propto\ell^{-2.44}$ \citep[Table 11,
$f_\mathrm{sky}^\mathrm{eff}=0.73$]{planckfg} and are highly anisotropic at
large scales, with their minimum in frequency space falling around 70--90 GHz
\citep[Fig. 51]{2016A&A...588A..65K, planckfg}.
This contamination can be mitigated by making
high signal-to-noise measurements of the CMB at degree scales and cleaning
foregrounds in multipole space,
which is the strategy of the ACTPol \citep{actpol}, BICEP \citep{bicep/keck},
 \textsc{Polarbear} \citep{polarbear}, and SPTPol \citep{sptpol}  experiments.
Another approach is to focus on large scale ($\theta\gtrsim10^\circ$)
fluctuations where it is computationally simpler to remove spatially varying foregrounds in map space,
an approach that has been employed using maps smoothed to $\theta\sim15^\circ$
\citep{wmapfinal,
plancklike15}.  For power spectrum-based analyses, the incomplete sky causes
issues both due to $\mathrm{E\to B}$ mixing (caused by spherical harmonics  no
longer forming a complete orthonormal basis) and the related issue that estimates of the CMB power spectrum $\hat C_\ell$ are not drawn from a well-understood  statistical distribution. 
These issues have been addressed by using $C_\ell$ estimators that
can reduce or specifically forbid $\mathrm{E\to B}$ mixing \citep[respectively]{polspice,purepseudocl}, and the
development of approximate likelihoods
that include any potential mixing effects explicitly
\citep{wishart, cross_cl}.

\citet{lowellpix} demonstrated
that the CLASS experiment, and
other experiments with multifrequency data and large observing area (e.g. 
LSPE, \citealt{lspe}, GroundBird, \citealt{groundbird} and PIPER, \citealt{gandilo}), will be able to overcome partial-sky $\mathrm{E\to
B}$-mode mixing and known sources of foreground contamination by using an
exact pixel-based likelihood for low-resolution measurements and a
pseudo-$C_\ell$ likelihood for higher-resolution measurements. In this paper, we
address mode mixing by fitting the model to the data using a pseudo-$C_\ell$ estimate from \texttt{PolSpice} \citep{polspice} and fitting the data to theory using the approximate Wishart distribution described in \citet{wishart}.

Another major obstacle to characterizing large angular scales is mitigating
systematic effects due to observations made on long timescales due to instrumental variations. 
To reach the necessary
instrumental stability, a front-end modulator in the form of a variable-delay
polarization modulator \citep[VPM,][]{Chuss:12} is used as the first optical element of each CLASS
telescope \citep{class}. This reduces instrumental effects well below the amplitude of an
$r=0.01$ signal \citep{miller}.

This paper expands on \citet{lowellpix} by characterizing the estimated power
spectrum across the entire angular range ($2\leqslant\ell\leqslant100$)  while simultaneously constraining
$\tau$, $A_s$,  $r$, and foreground emission, assuming  $1/f$ noise reduction to $r\ll0.01$ levels using a VPM \citep{miller}.  In addition to
quantifying the expected cosmological parameter constraints from the full CLASS
dataset, we also discuss constraints using combinations from external datasets.
CLASS will make a sample variance limited measurement of E-modes on the largest angular scales. With this precise measurement of  
$\tau$ ($\sigma_\tau\sim0.003$), the CLASS experiment's
measurements will  break the
$A_s\e^{-2\tau}$ partial degeneracy found in temperature anisotropy measurements. 
The resulting improved constraint on $A_s$ enables tighter bounds on the sum of neutrino masses $\sum m_\nu$.

In \autoref{sec:simulations} we will discuss our simulated data and the
assumptions that go into our modeling. \autoref{sec:likelihood} introduces our
implementation of the \citet{wishart} pseudo-$C_\ell$ likelihood and its efficacy at providing constraints given the
simulated data. \autoref{sec:externaldata} discusses the implications of a CLASS
$\tau$ measurement in the context of external cosmological parameter
constraints. Unless noted otherwise, all cosmological parameters are those
listed in \citet{plancktau}, specifically \texttt{PlanckTTTEEE+SIMlow}.

\section{Simulated Maps}
\label{sec:simulations}

\begin{figure*}
\centering
\includegraphics{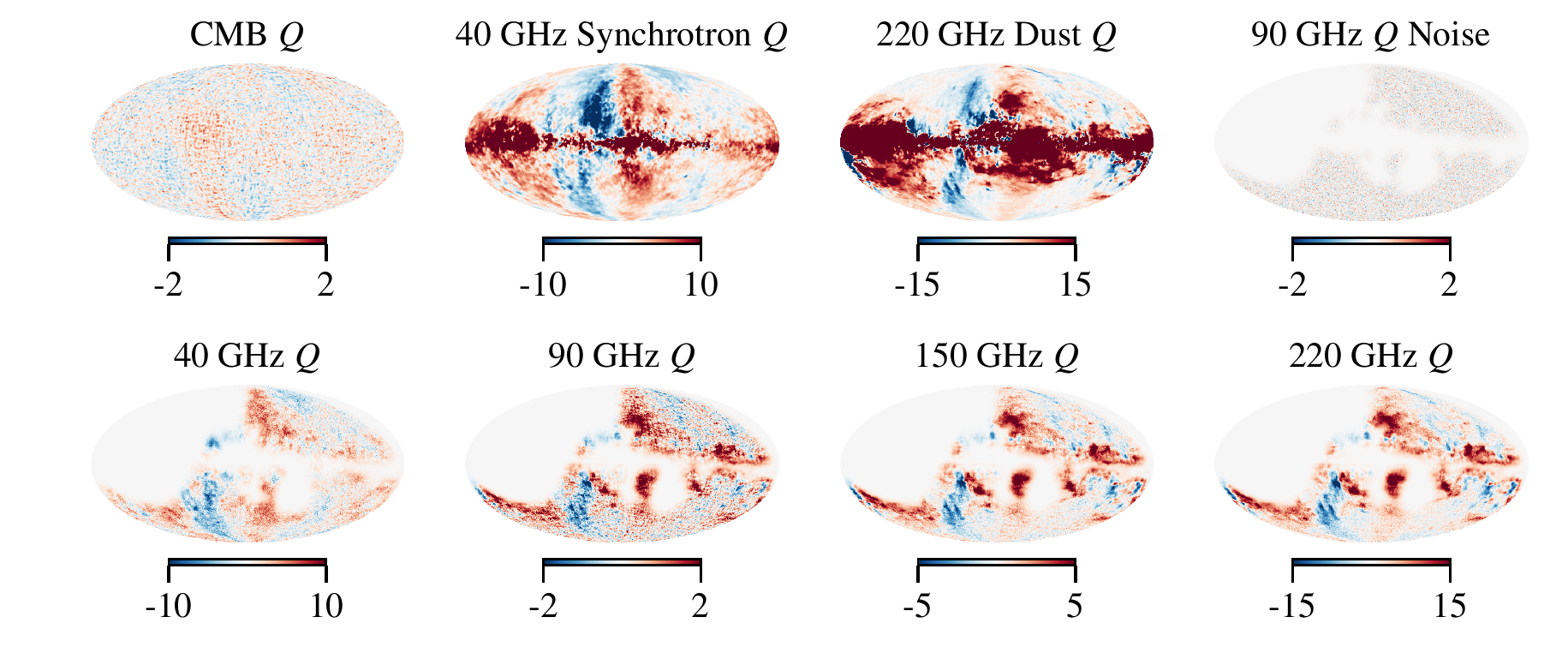}
\caption{
    Simulated CLASS maps include a realization of the CMB, polarized
synchrotron and thermal dust emission, and Gaussian white noise.
The top panels show the individual Stokes $Q$ components of the simulation, while the
bottom show the simulated multifrequency Stokes $Q$ CLASS maps, with the Galactic plane masked  and parts of the 
celestial Northern Hemisphere and celestial Southern Hemisphere excluded by the survey boundary. 
All maps are displayed in Galactic coordinates with units of $\mathrm{\mu K}$.
}
\label{fig:maps}
\end{figure*}

We use the CLASS instrument and survey specifications for our simulated
data as enumerated in \citet{class_spie}.
The CLASS experiment is located in the Atacama Desert in Chile, at a latitude of
$-23^\circ$, scanning 70\% of the sky every day at $45^\circ$ elevation.
We combine a mask due to the survey geometry with the \wmap\ P06 Galactic
foreground mask, which cuts out the brightest 25\% of the sky \citep{pol_mask}.
This leaves CLASS with an observed sky fraction of $f_\mathrm{sky}=0.47$.
The CLASS frequency bands are chosen to minimize
atmospheric emission
while straddling the Galactic foreground minimum.
Assuming a 5 year survey with 40, 90, 150, and 220 GHz channels, the maps are
assigned weights per pixel $w_{p,\nu}$, corresponding to white noise levels
$w_{p,\nu}^{-1/2}=[39,10,15,43]\ukarcmin$. We use this to simulate maps of
white noise as draws of a Gaussian random variable ${\boldsymbol n^\nu\sim\mathcal N(\boldsymbol 0,\sigma_\nu^2\mathbf
I)}$ with ${\sigma_\nu=w_{p,\nu}^{-1/2}/\sqrt{\Omega_\mathrm{pix}}}$, where
$\Omega_\mathrm{pix}$ is the area of a \texttt{HEALPix} pixel at the
simulated resolution, here $N_\mathrm{side}=128$\footnote{\texttt{HEALPix} \citep{healpix} maps
are divided into $12N_\mathrm{side}^2$ pixels, with each pixel width
$\theta_\mathrm{pix}\sim58.6^\circ/N_\mathrm{side}$. The full documentation can
be found at \url{http://healpix.sourceforge.net}.}.

We simulate foreground emission using 
\texttt{PySM}
\citep{pysm},\footnote{\url{https://github.com/bthorne93/PySM_public}} which
takes into account polarized foreground measurements from
\planck\ and \wmap\ 
(polarized dust from \citealt{planckfg}, polarized synchrotron from \citealt{wmapfinal}).
While it is known that the emission laws of these foregrounds vary
across the sky, with antenna temperature emission parametrized as
\begin{equation}
\label{eq:foregrounds}
\begin{aligned}
    \boldsymbol m^\nu_\mathrm{sync}&=\boldsymbol m_\mathrm{sync}
    \left(\frac{\nu}{\nu_\mathrm S}\right)^{\beta_\mathrm S(\nhat)}
    \\
    \boldsymbol m^\nu_\mathrm{dust}&=\boldsymbol
    m_\mathrm{dust}\left(\frac{\nu}{\nu_\mathrm D}\right)^{\beta_\mathrm D(\nhat)-2}
    \frac{B_\nu\big[T_\mathrm D(\nhat)\big]}
    {B_{\nu_\mathrm D}\big[T_\mathrm D(\nhat)\big]},
\end{aligned}
\end{equation}
current data do not yet meaningfully constrain the spatial variation of spectral indices within our sky cut
  (\citealt{lowellpix} Appendix B, \citealt{sheehy}, \citealt{newplanckdust}).
Therefore, we model foreground emission with fixed (i.e., isotropic) synchrotron spectral index
$\beta_\mathrm S$, dust spectral index $\beta_\mathrm D$, and blackbody emission $B_\nu[T_\mathrm D]$ with dust
temperature $T_\mathrm D$. Here we use $\nu_\mathrm S=40\ghz$ and
$\nu_\mathrm D=220\ghz$ as the reference frequencies, with $\boldsymbol
m_\mathrm{sync}$ and $\boldsymbol m_\mathrm{dust}$ the synchrotron and dust
emission at these respective frequencies.
The typical levels for these parameters are
$\beta_\mathrm S\sim-3.0\pm0.1$ \citep[\wmap\ intensity measurements]{specvar}, $\beta_\mathrm D\sim1.6\pm0.1$
\citep[\planck\ polarization measurements]{planckL2016}, and $T_\mathrm D\sim22\pm8 \kelvin$ \citep[\planck\ intensity
measurements]{planckfg}. While varying foreground emission laws are a
significant source of bias for B-mode measurements, E-modes are much brighter
and are largely unaffected by this source of uncertainty. Additionally, we
addressed this complication in \citet{lowellpix} by splitting the sky up into
subregions with constant emission parameters and showed that a 95\% C.L.~
measurement of $r=0.01$ was still possible. 
We have performed several simulations using the levels of spectral index
variation in \citet{pysm} ($\Delta\beta_\mathrm D<0.1$, $\Delta\beta_\mathrm S\sim0.1$) and found shifts in the recovery of $\tau$ on the order of
$\lesssim0.5\sigma_\tau$. These simulations used a single set of foreground maps that assumed instrumental white noise. 
For this work, we use $\beta_\mathrm D=1.6$ and
$\beta_\mathrm S=-3$ fixed across the sky.

For the CMB signal, we use the \texttt{CAMB} package
\citep{camb}\footnote{\url{http://camb.info}} to generate theoretical
$C_\ell^\mathrm{EE}$ and $C_\ell^\mathrm{BB}$, keeping all parameters fixed to
the \texttt{PlanckTTTEEE+SIMlow} \citet{plancktau} parameters, namely $\tau=0.0596$ and
$\ln(10^{10}A_s)=3.056$, with the addition of tensor B-modes of amplitude
$r=0.05$. With these theoretical power spectra in hand, we simulate maps using
\texttt{HEALPix}'s \texttt{synfast} function, from which we take the output $Q$
and $U$ Stokes parameters, denoted by the vector $\boldsymbol m_\mathrm{CMB}$.

The CLASS 40, 90, 150, and $220 \ghz$ bands have beam full width half maxima (FWHMs) of 90, 40, 24, and 18
arcmin, respectively, but for the purposes of this study we simulate the maps
with a common resolution of $1.5^\circ$.
We bring
all of the foregrounds and CMB to this common resolution
${\theta_\mathrm{FWHM}=1.5^\circ}$ and model the Gaussian noise as uncorrelated between pixels.

The data are from our multifrequency simulations 
\begin{equation}
    \label{eq:simulation}
\boldsymbol m^\nu=g(\nu)
\boldsymbol m_\mathrm{sync}^\nu + g(\nu)
\boldsymbol m_\mathrm{dust}^\nu
+\boldsymbol m_\mathrm{CMB} +
\boldsymbol n^\nu,
\end{equation}
where $g(\nu)\equiv\partial T/\partial T_A=(\e^x-1)^2/(x^2\e^x)$ is the conversion
factor from antenna to thermodynamic temperature referenced to the CMB radiation, 
with $x\equiv h\nu/kT_\mathrm{CMB}=\nu/(56.78\ghz)$, and $\boldsymbol m_{\mathrm{sync/dust}}^\nu$ as defined in \autoref{eq:foregrounds}. A single realization of the CLASS
Stokes $Q$ maps using this prescription is shown in \autoref{fig:maps}.

\section{Analysis Techniques}
\label{sec:likelihood}

\begin{figure*}
\centering
\includegraphics[width=\textwidth]{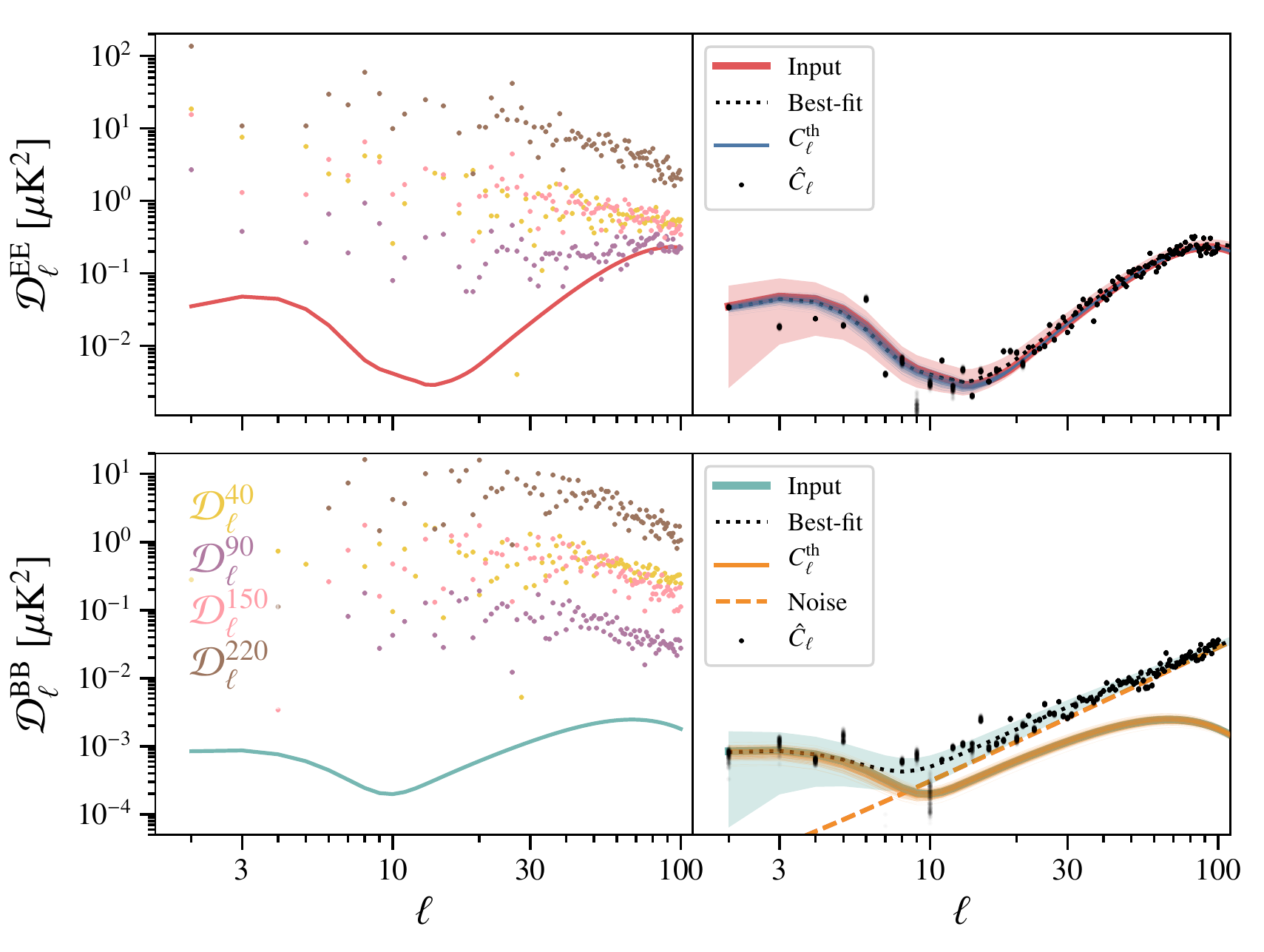}
\caption{Left: the points labeled $\mathcal D_\ell^\nu\equiv \ell(\ell+1)C_\ell^\nu/2\pi$ are the
autospectra associated with each CLASS band.
The amplitudes of these spectra depend both on the foreground amplitudes and the inherent noise bias in autospectra. In our likelihood, we include these autospectra and the cross-spectra (not plotted) and recover the input model (solid lines) by taking linear combinations of these 16 spectra. Note that the theoretical curves have been smoothed with a $1.5^\circ$ Gaussian window function. The B-mode spectrum includes contributions from primordial gravitational waves and gravitational lensing, the latter of which is subdominant for our fiducial value of $r=0.05$, but is dominant at the recombination peak when $r\lesssim0.01$.
    Right: this is a representation of the constraining power of CLASS in
    pseudo-$C_\ell$ space with each point and line being an independent draw from the MCMC chain.  The transparent overlapping gray dots ($\hat C_\ell$) represent estimates of the
    best-fit foreground-cleaned power spectrum
    ${\sum_{\nu_1,\nu_2}c_{\nu_1}c_{\nu_2}\hat C_\ell^{\nu_1\times\nu_2}}$ (with darker dots being many overlapping gray dots),
    the thin lines ($C_\ell^\mathrm{th}$) represent theory curves that were drawn from the chain, and
the thick solid lines (Input) the input theory power spectra. The white noise level (Noise) is plotted as an orange dashed line, and the best-fit $\mathrm{Theory+Noise}$ power spectrum is plotted as the black dotted line. The expected error
is represented by the transparent red and blue swaths, and is given in terms of
the input theory spectrum and the best-fit noise,
$\sigma_\ell=\sqrt{\frac2{(2\ell+1)f_\mathrm{sky}}}[C_\ell+N_\ell(c_\nu)]$. The input $r=0.05$, $\log 10^{10}A_s=3.046$, and $\tau=0.0596$, are all recovered within 95\% confidence levels.
}
\label{fig:power_spectrum}
\end{figure*}

\begin{figure*}
    \centering
    \includegraphics[width=\textwidth]{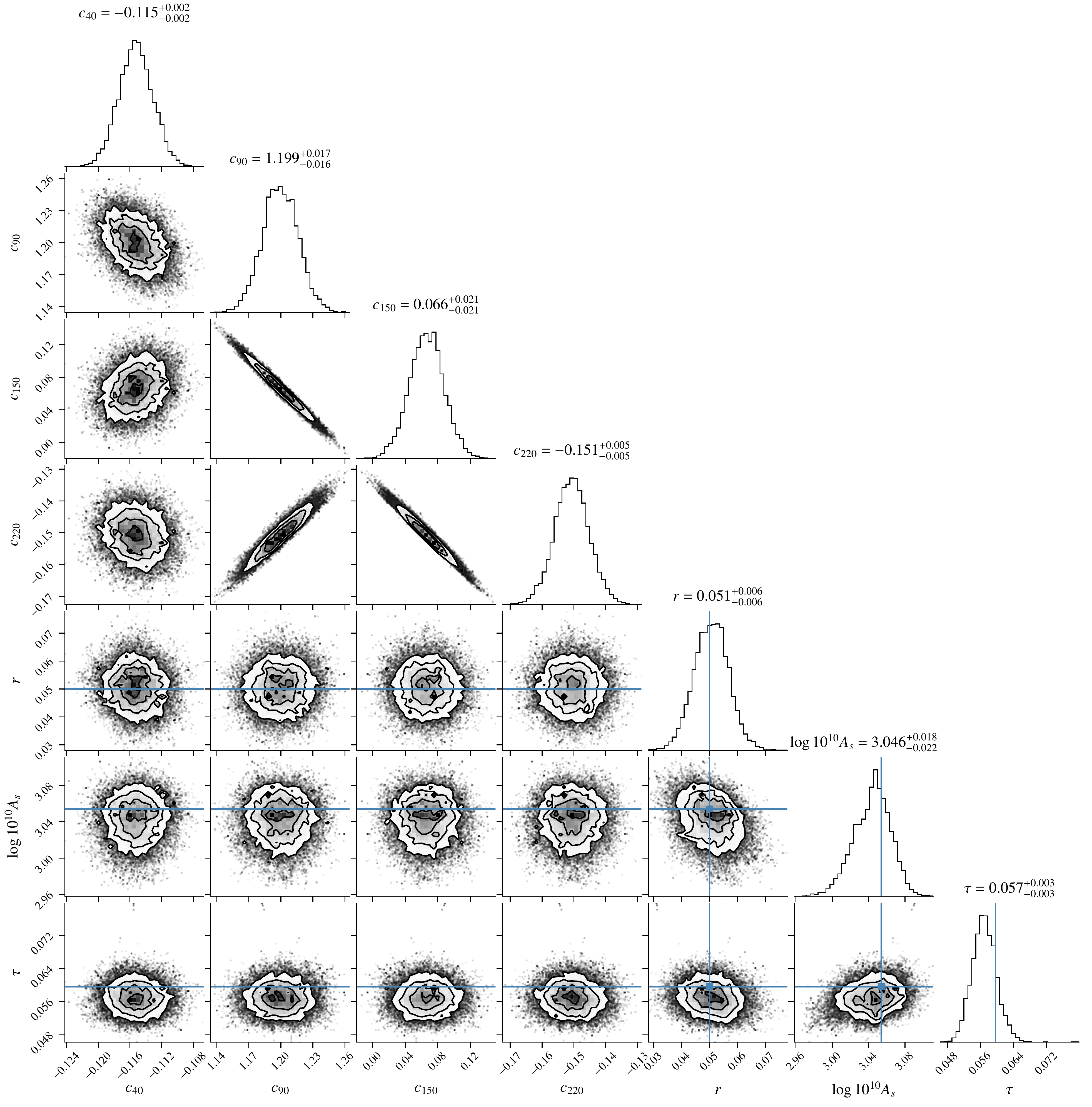}
\caption{Monte Carlo Chain for a single simulation. 
All parameter fluctuations are representative of the spread found in our suite of simulations.
The $\sim1\sigma$ offset in $\tau$ is not unexpected for this single realization. An accurate and unbiased $\tau$ results from many simulations.
The medians and their asymmetric 68\% confidence levels are quoted above each one-dimensional histogram.
The cosmological parameters are uncorrelated with the linear combination coefficients $c_\nu$, implying that any residual foregrounds do not affect parameter constraints. 
}
\label{fig:big_cornerplot}
\end{figure*}

For the CLASS experiment, E-modes are far
into the signal-dominated regime, with the main impediment to CMB
characterization being the Galactic foreground emission (assuming all
systematic measurement errors are under control). To estimate the linearly polarized Stokes parameters of the CMB maps and their polarized power spectra $C_\ell^\mathrm{EE}$ and $C_\ell^\mathrm{BB}$, we take linear combinations of the multifrequency maps constrained to keep the CMB amplitude consistent with blackbody emission,
\begin{equation}
    \label{eq:estimate}
\hat{\boldsymbol m}_\mathrm{CMB}\equiv
\sum_\nu c_\nu\boldsymbol m^\nu,\qquad\sum_\nu c_\nu=1.
\end{equation}
We ensure that the coefficients $c_\nu$ reduce foregrounds by imposing Gaussian priors
\begin{align}
\sum_\nu c_\nu g(\nu)\left(\frac \nu{\nu_\mathrm{S}}\right)^{\beta_\mathrm{S}}&=0\pm 0.01
\\
\sum_\nu c_\nu g(\nu)\left(\frac \nu{\nu_\mathrm{D}}\right)^{\beta_\mathrm{D}-2}
\frac{B_\nu\big[T_\mathrm D\big]}{B_{\nu_\mathrm D}\big[T_\mathrm D\big]}
&=0\pm 0.01,
\end{align}
corresponding to priors on $\Delta\beta_\mathrm{S/D}<0.1$.
This prior downweights unphysical solutions corresponding to values of $\beta_\mathrm{S/D}$ that are ruled out by existing data. This prior is relatively weak compared to the constraining power of the experiment, which returns constraints corresponding to $\Delta\beta_\mathrm S=0.02$ and $\Delta\beta_\mathrm D=0.005$.
If the $c_\nu$ are chosen such that there are no foreground
residuals while the instrumental noise contribution is minimized, the resulting
power spectrum estimate will be given by
\begin{equation}
\hat C_\ell^\mathrm{EE/BB}
=C_\ell^\mathrm{EE/BB}+\sum_\nu c_\nu^2N_\ell^\nu.
\end{equation}
where $N_\ell^\nu=w_{p,\nu}^{-1}$, in units of $\mu\mathrm
K^2\,\mathrm{sr}$. 

For our purposes, the foreground coefficients $c_\nu$ are nuisance parameters
that are marginalized over, while the true parameters of interest are
$r$, $\tau$, and $A_s$. To account for any spurious correlations between foregrounds
and CMB fluctuations,  we simultaneously fit for the foreground coefficients and
the cosmological parameters.
Given the power spectrum estimate $\hat C_\ell(c_\nu)$, the noise power spectrum
$N_\ell=\sum_\nu c_\nu^2N_\ell^\nu$, and the theoretical power spectrum
$C_\ell(r,A_s,\tau)$, the cut-sky likelihood for the power spectra
$C_\ell^\mathrm{EE/BB}$ is given by minimizing
\begin{gather}
\begin{aligned}
    \label{eq:likeapprox}
    -2\ln\mathcal L\simeq&\sum_{\ell\ell'}\left[G\left(\frac{\hat
    C_\ell}{C_\ell+N_\ell}\right)C_{f\ell}\right]
    [M_f^{-1}]_{\ell\ell'}\left[C_{f\ell'}G\left(\frac{\hat
    C_{\ell'}}{C_{\ell'}+N_{\ell'}}\right)\right]
    \\
    &+2\sum_\ell\ln|\hat C_\ell|
\end{aligned}
\end{gather}
where $G(x)\equiv\sqrt{2(x-\ln x-1)}$. The subscript $f$ refers to some fiducial
model, and $\boldsymbol M_f$ is the covariance of $\hat C_\ell$ evaluated for
the fiducial model $C_{f\ell}$ \citep[Equation 50 of][see
\autoref{sec:determinant} for an explanation of the final term]{wishart}. 
We split the covariance matrix into two terms, one with CMB and white noise, and another with foreground residuals, $\boldsymbol M_f\equiv\boldsymbol M_f^\mathrm{C+N}+\boldsymbol M_f^\mathrm{fore}$.
We estimate
$\boldsymbol M_f^\mathrm{C+N}$ using simulated data on a cut sky with only CMB and Gaussian white noise contributions,
using $r=0.05$, $\ln(10^{10}A_s)=3.056$,  $\tau=0.0596$, and $w_p^{-1/2}=14\,\mathrm{\mu K\,arcmin}$ as the fiducial
model parameters. The estimated covariance matrix has
$M^\mathrm{C+N}_{f,(\ell,\ell+1)}/M^\mathrm{C+N}_{f,(\ell,\ell)}\lesssim0.1$, with most values $<0.03$ at the 68\% C.L. The diagonal elements
$M^\mathrm{C+N}_{f,\ell\ell}$ agree with the analytical prediction from \citet{polspice} at the 5\% level,
\begin{equation}
M^\mathrm{C+N}_{f,(\ell\ell)}
=\frac2{(2\ell+1)f_\mathrm{sky}w_2^2/w_4}C_\ell^2,
\end{equation}
where $w_n=\int w(\nhat)^n\ud\Omega$, $w(\nhat)$ is the apodized mask, and $f_\mathrm{sky}=w_1$ is the observed sky fraction. 

The addition of a term $\boldsymbol M^\mathrm{fore}_f$ accounts for any foreground residuals encountered during the fits. In principle, the best-fit solution does  not have any foreground contribution, but any variation around this point in parameter space will affect the best-fit value and will potentially induce spurious correlations. To estimate the effect of foreground residuals, we took the $c_\nu$ of a successful MCMC chain without any foreground covariance accounted for and computed $\hat C_\ell$ of foreground residuals taken from multifrequency maps without noise or CMB. This gave a sample covariance matrix $\boldsymbol M_f^\mathrm{fore}$. Using this, we recomputed the Monte Carlo chain using this extra covariance, and found that the recovered cosmological parameters were accurately reconstructed, with an increase in their uncertainty, e.g. for the chain used in \autoref{fig:power_spectrum}, ${\sigma_r=0.0048\to0.0064}$ and $\sigma_\tau=0.0022\to0.0029$.

%

%

We estimate the pseudo-$C_\ell$ power spectrum using \texttt{PolSpice} \citep{polspice}, which
corrects for the effects of masking and inter-bin correlations induced by the
incomplete sky. We represent the
estimation of the power spectrum using a bilinear operator $\mathsf P$ such that
$\hat C_\ell=\boldsymbol m^T\mathsf P\boldsymbol m$. In practice, we use the
bilinear property of this operator to take sums of all multifrequency cross-spectra and subtract foregrounds in multipole space, i.e.,
\begin{equation}
\hat C_\ell=
\left(\sum_{\nu_1}c_{\nu_1}\boldsymbol m^{\nu_1}\right)^T
\mathsf P
\left(\sum_{\nu_2}c_{\nu_2}\boldsymbol m^{\nu_2}\right)
=\sum_{\nu_1,\nu_2}c_{\nu_1}c_{\nu_2}\hat C_\ell^{\nu_1\times\nu_2}
\end{equation}
where we have defined $\hat C_\ell^{a\times b}\equiv (\boldsymbol m^a)^T\mathsf P\boldsymbol m^b$.

The method outlined here reduces and accounts for any $\mathrm{E\to B}$ mixing inherent in the analysis of an incomplete sky while accounting for the underlying statistical distribution of the power spectrum. \texttt{PolSpice} returns a decoupled estimate of the polarization power spectra, giving an unbiased estimate of the true underlying power spectrum while minimizing spurious correlations between E-modes and B-modes. The approximate Wishart distribution from \citet{wishart} accounts for the non-Gaussian nature of the low-$\ell$ power spectra while explicitly accounting for any residual E-B correlation in the fiducial covariance matrix $\boldsymbol M_f$.

Because calls to \texttt{CAMB} are computationally expensive, with each call taking $\mathcal
O(1\,\mathrm{sec})$, we have written a code
\texttt{clee-fast}\footnote{\url{https://github.com/pqrs6/clee-fast} \citep{cleefast}} that
linearly interpolates between precomputed power spectra, only allowing
variation in $r$, $A_s$, and $\tau$. This is similar in spirit to
\texttt{PICO} \citep{pico}, but works better for our purposes because it only
allows variation of three parameters, reducing numerical noise and computational cost. Examples of the approximated theory curves and
pseudo-$C_\ell$ estimates are displayed in 
\autoref{fig:power_spectrum}, and the corresponding corner plot of the parameter chain is displayed in \autoref{fig:big_cornerplot}.

\section{Predicting parameter constraints}
\label{sec:externaldata}

\begin{figure}
\begin{center}
    \includegraphics[width=0.5\textwidth]{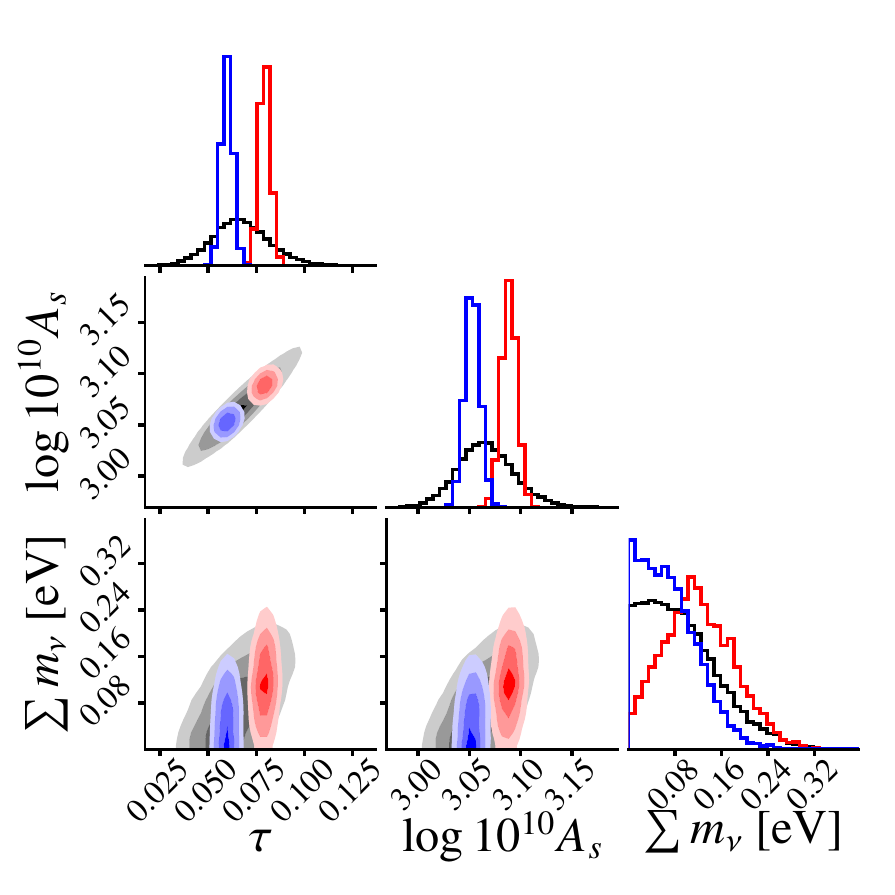}
\end{center}
    \caption{Combination with the \planck\ MCMC chains
        \texttt{base\_mnu\_plikHM\_TTTEEE\_lowTEB\_lensing\_BAO}, with CLASS
    posteriors applied using $\tau=0.060\pm0.003$ (blue) and $\tau=0.080\pm0.003$ (red).
These two cases give neutrino mass
constraints $\sum m_\nu=64^{+56}_{-45}\mev$ (blue) and $\sum m_\nu=117\pm60\mev$ (red). The black contours are  from the raw \planck\ chains, and yield constraints $\tau=0.067^{+0.015}_{-0.014}$ and $\sum m_\nu=88^{+73}_{-55}\mev$.
}
    \label{fig:mnu_contours}
\end{figure}

We obtain sample variance limited constraints that are on the order of
$\sigma_\tau/\tau\sim 5\%$.
This
is a factor of $\sim3$ improvement on the \planck\ precision of  $\sigma_\tau/\tau\sim16\%$, and is a
factor of two away from the full-sky cosmic variance precision,
$\sigma_\tau/\tau\sim 2.5\%$. We note that there exists a publicly available code, \texttt{cmb4cast} \citep{cmb4cast}, that uses Fisher matrix analyses to make similar projections. This code gives $\sigma_\tau=0.0035$, slightly larger than our $\sigma_\tau=0.0029$. This discrepancy comes from a number of different assumptions between the codes, such as the level of foreground variation, priors on foreground variation, and the fiducial cosmological parameters. Despite these differences, it is reassuring that these different approaches yield this level of agreement.

While large-scale polarization measurements are weakly sensitive to variations
in $A_s$, the strong E-mode sensitivity to $\tau$ can break the
partial degeneracy in the well-constrained parameter combination $A_s\e^{-2\tau}$ found
in intensity measurements.
The amplitude of primordial scalar fluctuations $A_s$ can be used to predict the
amplitude of matter fluctuations at low redshifts in the linear regime,
typically parameterized by the amplitude of dark matter density fluctuations at a scale of
$8\,h^{-1}\mpc$, $\sigma_8$. 

In standard \lcdm, there are three neutrino
species,
and there is experimental evidence that there is a nonzero difference in the
squares of each neutrino species' mass, which is detected via the oscillation of
neutrinos from one species to another as they travel through space \citep{lsnd, cmbneutrinos}. 
In the
normal hierarchy, the mass of one neutrino is much greater than the other two, which requires that the sum of
the neutrino masses $\sum m_\nu>60\mev$. In the inverted hierarchy, two neutrinos have similar masses that are much larger than the third, which requires $\sum m_\nu>100\mev$ \citep[Section 14.2]{pdg17}.

In the early universe, before neutrinos became nonrelativistic matter, massive neutrinos at small scales free streamed, effectively
reducing the amplitude of matter fluctuations \citep{collisionless_damping,analytic_damping}. 
In this way the neutrino mass affects the cosmological model's prediction for $\sigma_8$ given $A_s$.
This effect can be used to
constrain the mass of neutrinos from above, with current upper limits $\sum
m_\nu<170\mev$ at the 95\% C.L. using \planck\ temperature and low-$\ell$ polarization measurements, 
in combination with BOSS DR12 BAO data and the JLA Type Ia SNe catalog
\citep{couchot_neutrino}. 
Tighter constraints on $\sigma_8$ should improve these limits, although the
latest results from the Dark Energy Survey \citep[DES,][]{des1yr} using galaxy clustering and weak lensing
show that adding these data to the \planck+JLA+BAO data actually increases the
95\% C.L.~by 20\%, which can be attributed to the tension in the values of $\sigma_8$
inferred by \planck\ and DES.
There is enough uncertainty in $A_s$,
mainly due to the partial degeneracy in $A_s\e^{-2\tau}$, 
to weaken any $\sum m_\nu$ measurement to the $\sim2\sigma$ level
for the minimal $\sum m_\nu =60 \mev$ scenario allowed by neutrino oscillations.

\citet{allison} use Fisher forecasts of future measurements to predict the constraints from combining
low-$\ell$ polarization measurements
with the Dark Energy Spectroscopic Instrument (DESI) and CMB-S4.
In particular, using 
$\ell_\mathrm{min}=50$ for CMB-S4 yields
$\sigma_{\sum m_\nu}\sim27\mev$ with only \wmap\ low-$\ell$ polarization data, and $19\mev$, using pre-2016 
\planck\ low-$\ell$ polarization sensitivities. 
These upper limits are inflated by uncertainty in $A_s$ from the partial degeneracy with $\tau$. Therefore, 
an external constraint on $\tau$ can break this
degeneracy, allowing for any differences between the $A_s$ prediction of
$\sigma_8$ and the measured value of $\sigma_8$ to be directly and precisely computed.
In the case
of putative CMB-S4 measurements with $\ell_\mathrm{min}=5$, $\sigma_{\sum m_\nu}$ is reduced to $15\mev$, with the
reduction in uncertainty coming almost entirely from the uncertainty on $\tau$
reducing to $\sigma_\tau=0.003$. 
As we have shown, if CLASS is able to measure $C_\ell^\mathrm{EE}$ and $C_\ell^\mathrm{BB}$ down to $\ell=2$ with white noise, it will achieve this $\sigma_\tau$.
In \autoref{fig:mnu_contours}, using \planck\ MCMC chains from the 2015 data release,\footnote{\texttt{COM\_CosmoParams\_fullGrid\_R2.00.tar.gz} from the \planck\ Legacy Archive \href{https://pla.esac.esa.int/pla/}{https://pla.esac.esa.int/pla/}}
we show 
how a CLASS $\tau$ measurement would improve constraints on
$\sigma_{\sum m_\nu}$ with currently available data.

\section{Conclusions}
\label{sec:conclusions}

We have implemented a $\hat C_\ell$-based likelihood for
large-scale polarized CMB measurements in the presence of polarized foregrounds
and instrumental noise measured on a partial sky.
To do this, we implemented a fast interpolation scheme for retrieving $C_\ell(r,A_\mathrm s,\tau)$, and used \texttt{PolSpice} to develop a pseudo-$C_\ell$ likelihood that takes into account mode coupling from a cut-sky analysis.

\begin{enumerate}
    \item We recover the input reionization optical depth with
        $\sigma_\tau\sim0.003$, within a factor of two of the cosmic variance
        limited case.
    \item We recover the tensor-to-scalar ratio with $\sigma_r\sim0.006$,
        consistent with our partial pixel-based method in \citet{lowellpix}.
    \item We demonstrate the power of a $\tau$ prior on massive neutrino
        constraints $\sum m_\nu$ using \planck{} Monte Carlo chains.
\end{enumerate}

The CLASS experiment was designed to characterize the large-scale polarized CMB up to a sensitivity that allows a $2\sigma$ measurement of primordial gravitational waves with an amplitude of $r=0.01$. As we have demonstrated, satisfying this requirement by measuring $C_\ell^\mathrm{EE}$ and $C_\ell^\mathrm{BB}$ down to $\ell=2$ necessarily yields an estimate of the reionization optical depth $\tau$ that is limited only by sample variance and cannot be meaningfully improved upon using measurements of the CMB alone. 
CLASS's $\tau$ constraint will be critical in characterizing 
neutrino mass,
helping to fulfill a major objective in both the particle physics and cosmology communities.

\appendix

\section{Derivation of pseudo-$C_\ell$ likelihood}
\label{sec:determinant}

In \citet{wishart}, the cut-sky approximation for a pseudo-$C_\ell$ likelihood 
is derived assuming a fixed power spectrum estimate. However, by the nature of the 
joint cosmological parameter and foreground cleaning estimates, this is not
accurate for our purposes, and the best-fit solution ends up being one where the
total noise level is increased without bound. Here we review the calculations of
\citeauthor{wishart} while keeping the dependence of the estimated $\hat C_\ell$ explicit.

The spherical harmonic coefficient vector $\mathbf a_{\ell m}=(a_{\ell m}^\mathrm T,a_{\ell m}^\mathrm
E,a_{\ell m}^\mathrm B)^T$ is a normally distributed random variable with covariance matrix at each $\ell$
\begin{equation}
    \boldsymbol C_\ell\equiv\langle\mathbf a_{\ell m}^{\phantom\dagger}
    \mathbf a_{\ell
m}^\dagger\rangle
\end{equation}
and estimator
\begin{equation}
    \hat{\boldsymbol C}_\ell\equiv\frac1{2\ell+1}\sum_m\mathbf a_{\ell
    m}^{\phantom\dagger}
    \mathbf
    a_{\ell m}^\dagger.
\end{equation}
In standard \lcdm, the $\mathbf a_{\ell m}$ are Gaussian distributed with
\begin{equation}
    -2\ln P(\{\mathbf a_{\ell m}\}|\boldsymbol C_\ell)
    =\sum _m[\mathbf a_{\ell m}^\dagger\boldsymbol C_\ell^{-1}\mathbf a_{\ell
    m}^{\phantom\dagger}+\ln|2\pi\boldsymbol C_\ell|]
    =(2\ell+1)\big(\mathrm{Tr}[\hat{\boldsymbol C}_\ell\boldsymbol
    C_\ell^{-1}]+\ln|\boldsymbol C_\ell|\big)+\mathrm{const.}
\end{equation}
For a full-sky likelihood, $\hat{\boldsymbol C}_\ell$ contains all of the sky's
information, and is drawn from a Wishart distribution,
\begin{equation}
    P(\hat{\boldsymbol C}_\ell|\boldsymbol C_\ell)\propto
    \frac{|\hat{\boldsymbol C}_\ell|^{(2\ell-n)/2}}{|\boldsymbol
        C_\ell|^{(2\ell+1)/2}}\e^{-(2\ell+1)\mathrm{Tr}(\hat{\boldsymbol
        C}_\ell\boldsymbol C_\ell^{-1})/2}
\end{equation}
where $n$ is the number of fields considered. We take $n=2$ for 
$a_{\ell m}^\mathrm E$ and $a_{\ell m}^\mathrm B$.

The root of the \citeauthor{wishart} approximation involves rewriting this likelihood
in a quadratic form. Using orthogonal $\boldsymbol U_\ell$ and diagonal
$\boldsymbol{{D}}_\ell$ to rewrite $\boldsymbol C_\ell^{-1/2}\hat{\boldsymbol
C}_\ell{\boldsymbol C}_\ell^{-1/2}\equiv\boldsymbol U_\ell\boldsymbol
{D}_\ell\boldsymbol U_\ell^T$, the probability can be written as
\begin{align}
    -2\ln P&=(2\ell+1)\left\{\mathrm{Tr}[\hat{\boldsymbol C}_\ell\boldsymbol
    C_\ell^{-1}]+\ln|\boldsymbol
    C_\ell|-\frac{2\ell-n}{2\ell+1}\ln|\hat{\boldsymbol
    C}_\ell|\right\}+\mathrm{const.} \label{eq:approx}
    \\
    &=(2\ell+1)\left\{
        \mathrm{Tr}[\hat{\boldsymbol C}_\ell\boldsymbol
        C_\ell^{-1}]-\ln|\hat{\boldsymbol C}_\ell\boldsymbol C_\ell^{-1}|-n
    \right\}
    +n\Big[\ln|\hat{\boldsymbol C}_\ell|+2\ell+1\Big]+\mathrm{const.}
    \\
    &=\frac{2\ell+1}2\mathrm{Tr}[\boldsymbol G(\boldsymbol {D}_\ell)]^2
    +n\ln|\hat{\boldsymbol C}_\ell|+\mathrm{const.}
\end{align}
where $G(x)\equiv\sqrt{2(x-\ln x-1)}$ and $[\boldsymbol G(\boldsymbol
{D}_\ell)]_{ij}=G(D_{\ell,ii})\delta_{ij}$.

Note that, in \autoref{eq:approx}, if we assume that $\hat{\boldsymbol C}_\ell$
is constant, we can adjust the constant such that $-2\ln P=0$ when $\boldsymbol
C_\ell=\hat{\boldsymbol C}_\ell$. This is where our derivation differs from
\citeauthor{wishart}.
From here, the derivation in \citeauthor{wishart} applies, carrying along the extra
$n\ln|\hat{\boldsymbol C}_\ell|$ term. This term essentially adds a penalty for
increasing the noise, preventing the coefficients in $\hat{\boldsymbol
C}_\ell=\sum_{\nu_1,\nu_2}c_{\nu_1}c_{\nu_2}\hat{\boldsymbol
C}_\ell^{\nu_1\times\nu_2}$ from getting too large.

\acknowledgments

We acknowledge the National Science Foundation Division of Astronomical Sciences
for their support of CLASS under grant Nos. 0959349, 1429236, 1636634, and
1654494. 
The CLASS project employs detector technology developed under several previous and ongoing NASA grants.
Detector development work at JHU was funded by NASA grant number NNX14AB76A. 
K.H. is supported by the NASA Space Technology Research Fellowship grant number NXX14AM49H.
We
further acknowledge the very generous support of Jim and Heather Murren
(JHU~A\&S~'88), Matthew Polk (JHU~A\&S~Physics BS~'71), David Nicholson, and Michael
Bloomberg (JHU~Engineering~'64). CLASS is located in the Parque Astron\'omica
Atacama in northern Chile under the auspices of the Comisi\'on Nacional de
Investigaci\'on Cient\'\i fica y Tecnol\'ogica de Chile (CONICYT).
Some of the results in this paper have been derived using the \texttt{HEALPix}
\citep{healpix} package.
We also acknowledge use of the
\planck\ Legacy Archive. \planck\ is an ESA science mission with instruments and contributions directly funded
by ESA Member States, NASA, and Canada.
Part of this research project was conducted using computational resources at the Maryland Advanced Research Computing Center (MARCC).
D.J.W. thanks Graeme Addison and Janet Weiland for productive conversations, and Kirill Tchernyshyov for engaging in fruitful discussions regarding statistics. We also thank the anonymous referee for their helpful comments that improved our final manuscript.

\software{\texttt{python}, \texttt{IPython} \citep{ipython}, \texttt{numpy} \citep{numpy}, \texttt{scipy} \citep{scipy}, \texttt{matplotlib} \citep{matplotlib}, \texttt{healpy} \citep{healpix}, \texttt{mpi4py} \citep{mpi4py3}, \texttt{corner} \citep{corner}, \texttt{emcee} \citep{emcee}, \texttt{PolSpice} \citep{polspice}, \texttt{camb} \citep{camb}}
\bibliographystyle{aasjournal}
\bibliography{ms}

\end{document}

%% file: commands.tex
\newcommand{\ud}{\,\mathrm{d}}
\newcommand{\der}[3][{}]{\frac{\mathrm{d}^{#1}#2}{\mathrm{d}#3^{#1}}}
\newcommand{\pder}[3][{}]{\frac{\partial^{#1}#2}{\partial#3^{#1}}}
\newcommand{\dd}{\mathrm{d}}
\newcommand{\zhat}{\boldsymbol{\hat z}}
\newcommand{\xhat}{\boldsymbol{\hat x}}
\newcommand{\yhat}{\boldsymbol{\hat y}}
\newcommand{\nhat}{\boldsymbol{\hat n}}
\newcommand{\e}{e}
\newcommand{\lcdm}{$\Lambda$CDM}
\newcommand{\wmap}{\textit{WMAP}}
\newcommand{\planck}{\textit{Planck}}

\newcommand{\pc}{\,\mathrm{pc}}
\newcommand{\kpc}{\,\mathrm{kpc}}
\newcommand{\mpc}{\,\mathrm{Mpc}}
\newcommand{\gpc}{\,\mathrm{Gpc}}
\newcommand{\meter}{\,\mathrm{m}}
\newcommand{\cm}{\,\mathrm{cm}}
\newcommand{\km}{\,\mathrm{km}}
\newcommand{\nm}{\,\mathrm{nm}}
\newcommand{\mm}{\,\mathrm{mm}}
\newcommand{\au}{\,\mathrm{AU}}
\newcommand{\mum}{\,\mu\mathrm{m}}
\newcommand{\picom}{\,\mathrm{pm}}
\newcommand{\A}{\,\text{\AA}}

\newcommand{\rad}{\,\mathrm{rad}}

\newcommand{\ev}{\,\mathrm{eV}}
\newcommand{\kev}{\,\mathrm{keV}}
\newcommand{\Mev}{\,\mathrm{MeV}}
\newcommand{\mev}{\,\mathrm{meV}}
\newcommand{\Gev}{\,\mathrm{GeV}}
\newcommand{\Tev}{\,\mathrm{TeV}}
\newcommand{\joule}{\,\mathrm J}
\newcommand{\watts}{\,\mathrm W}
\newcommand{\erg}{\,\text{erg}}

\newcommand{\g}{\,\mathrm{g}}
\newcommand{\kg}{\,\mathrm{kg}}

\newcommand{\s}{\,\mathrm s}
\newcommand{\mus}{\,\mu\mathrm s}
\newcommand{\yr}{\,\mathrm{yr}}
\newcommand{\gyr}{\,\mathrm{Gyr}}

\newcommand{\ghz}{\,\mathrm{GHz}}
\newcommand{\hz}{\,\mathrm{Hz}}
\newcommand{\mhz}{\,\mathrm{MHz}}

\newcommand{\kelvin}{\,\mathrm K}
\newcommand{\uk}{\,\mu\mathrm K}
\newcommand{\sr}{\,\mathrm{sr}}

\newcommand{\muG}{\,\mu\mathrm G}

\newcommand{\ukarcmin}{\,\mu\mathrm{K\,arcmin}}

\def\sectionautorefname{Section}
\def\subsectionautorefname{Subsection}

%% file: ms.bbl
\begin{thebibliography}{}
\expandafter\ifx\csname natexlab\endcsname\relax\def\natexlab#1{#1}\fi
\providecommand{\url}[1]{\href{#1}{#1}}

\bibitem[{{Abazajian} {et~al.}(2015){Abazajian}, {Arnold}, {Austermann},
  {Benson}, {Bischoff}, {Bock}, {Bond}, {Borrill}, {Calabrese}, {Carlstrom},
  {Carvalho}, {Chang}, {Chiang}, {Church}, {Cooray}, {Crawford}, {Dawson},
  {Das}, {Devlin}, {Dobbs}, {Dodelson}, {Dor{\'e}}, {Dunkley}, {Errard},
  {Fraisse}, {Gallicchio}, {Halverson}, {Hanany}, {Hildebrandt}, {Hincks},
  {Hlozek}, {Holder}, {Holzapfel}, {Honscheid}, {Hu}, {Hubmayr}, {Irwin},
  {Jones}, {Kamionkowski}, {Keating}, {Keisler}, {Knox}, {Komatsu}, {Kovac},
  {Kuo}, {Lawrence}, {Lee}, {Leitch}, {Linder}, {Lubin}, {McMahon}, {Miller},
  {Newburgh}, {Niemack}, {Nguyen}, {Nguyen}, {Page}, {Pryke}, {Reichardt},
  {Ruhl}, {Sehgal}, {Seljak}, {Sievers}, {Silverstein}, {Slosar}, {Smith},
  {Spergel}, {Staggs}, {Stark}, {Stompor}, {Vieregg}, {Wang}, {Watson},
  {Wollack}, {Wu}, {Yoon}, \& {Zahn}}]{cmbneutrinos}
{Abazajian}, K.~N., {Arnold}, K., {Austermann}, J., {et~al.} 2015,
  Astroparticle Physics, 63, 66

\bibitem[{Abazajian {et~al.}(2016)Abazajian, Adshead, Ahmed, Allen, Alonso,
  Arnold, Baccigalupi, Bartlett, Battaglia, Benson, Bischoff, Borrill, Buza,
  Calabrese, Caldwell, Carlstrom, Chang, Crawford, Cyr-Racine, {De Bernardis},
  de~Haan, Alighieri, Dunkley, Dvorkin, Errard, Fabbian, Feeney, Ferraro,
  Filippini, Flauger, Fuller, Gluscevic, Green, Grin, Grohs, Henning, Hill,
  Hlozek, Holder, Holzapfel, Hu, Huffenberger, Keskitalo, Knox, Kosowsky,
  Kovac, Kovetz, Kuo, Kusaka, Jeune, Lee, Lilley, Loverde, Madhavacheril,
  Mantz, Marsh, McMahon, Meerburg, Meyers, Miller, Munoz, Nguyen, Niemack,
  Peloso, Peloton, Pogosian, Pryke, Raveri, Reichardt, Rocha, Rotti, Schaan,
  Schmittfull, Scott, Sehgal, Shandera, Sherwin, Smith, Sorbo, Starkman, Story,
  van Engelen, Vieira, Watson, Whitehorn, \& Wu}]{s4book}
Abazajian, K.~N., Adshead, P., Ahmed, Z., {et~al.} 2016, arXiv:1610.02743.
\newblock \url{http://arxiv.org/abs/1610.02743}

\bibitem[{{Addison} {et~al.}(2016){Addison}, {Huang}, {Watts}, {Bennett},
  {Halpern}, {Hinshaw}, \& {Weiland}}]{addison}
{Addison}, G.~E., {Huang}, Y., {Watts}, D.~J., {et~al.} 2016, \apj, 818, 132

\bibitem[{{Aiola} {et~al.}(2012){Aiola}, {Amico}, {Battaglia}, {Battistelli},
  {Ba{\'o}}, {de Bernardis}, {Bersanelli}, {Boscaleri}, {Cavaliere},
  {Coppolecchia}, {Cruciani}, {Cuttaia}, {D'Addabbo}, {D'Alessandro}, {De
  Gregori}, {Del Torto}, {De Petris}, {Fiorineschi}, {Franceschet},
  {Franceschi}, {Gervasi}, {Goldie}, {Gregorio}, {Haynes}, {Krachmalnicoff},
  {Lamagna}, {Maffei}, {Maino}, {Masi}, {Mennella}, {Morgante}, {Nati}, {Ng},
  {Pagano}, {Passerini}, {Peverini}, {Piacentini}, {Piccirillo}, {Pisano},
  {Ricciardi}, {Rissone}, {Romeo}, {Salatino}, {Sandri}, {Schillaci},
  {Stringhetti}, {Tartari}, {Tascone}, {Terenzi}, {Tomasi}, {Tommasi}, {Villa},
  {Virone}, {Withington}, {Zacchei}, \& {Zannoni}}]{lspe}
{Aiola}, S., {Amico}, G., {Battaglia}, P., {et~al.} 2012, in Society of
  Photo-Optical Instrumentation Engineers (SPIE) Conference Series, Vol. 8446,
  Society of Photo-Optical Instrumentation Engineers (SPIE) Conference Series

\bibitem[{{Albrecht} \& {Steinhardt}(1982)}]{albrecht}
{Albrecht}, A., \& {Steinhardt}, P.~J. 1982, Physical Review Letters, 48, 1220

\bibitem[{{Allison} {et~al.}(2015){Allison}, {Caucal}, {Calabrese}, {Dunkley},
  \& {Louis}}]{allison}
{Allison}, R., {Caucal}, P., {Calabrese}, E., {Dunkley}, J., \& {Louis}, T.
  2015, \prd, 92, 123535

\bibitem[{Athanassopoulos {et~al.}(1998)Athanassopoulos, Auerbach, Burman,
  Caldwell, Church, Cohen, Donahue, Fazely, Federspiel, Garvey, Gunasingha,
  Imlay, Johnston, Kim, Louis, Majkic, McIlhany, Mills, Reeder, Sandberg,
  Smith, Stancu, Strossman, Tayloe, VanDalen, Vernon, Wadia, Waltz, White,
  Works, Xiao, \& Yellin}]{lsnd}
Athanassopoulos, C., Auerbach, L.~B., Burman, R.~L., {et~al.} 1998, Phys. Rev.
  Lett., 81, 1774.
\newblock \url{https://link.aps.org/doi/10.1103/PhysRevLett.81.1774}

\bibitem[{{Austermann} {et~al.}(2012){Austermann}, {Aird}, {Beall}, {Becker},
  {Bender}, {Benson}, {Bleem}, {Britton}, {Carlstrom}, {Chang}, {Chiang},
  {Cho}, {Crawford}, {Crites}, {Datesman}, {de Haan}, {Dobbs}, {George},
  {Halverson}, {Harrington}, {Henning}, {Hilton}, {Holder}, {Holzapfel},
  {Hoover}, {Huang}, {Hubmayr}, {Irwin}, {Keisler}, {Kennedy}, {Knox}, {Lee},
  {Leitch}, {Li}, {Lueker}, {Marrone}, {McMahon}, {Mehl}, {Meyer}, {Montroy},
  {Natoli}, {Nibarger}, {Niemack}, {Novosad}, {Padin}, {Pryke}, {Reichardt},
  {Ruhl}, {Saliwanchik}, {Sayre}, {Schaffer}, {Shirokoff}, {Stark}, {Story},
  {Vanderlinde}, {Vieira}, {Wang}, {Williamson}, {Yefremenko}, {Yoon}, \&
  {Zahn}}]{sptpol}
{Austermann}, J.~E., {Aird}, K.~A., {Beall}, J.~A., {et~al.} 2012, in Society
  of Photo-Optical Instrumentation Engineers (SPIE) Conference Series, Vol.
  8452, Society of Photo-Optical Instrumentation Engineers (SPIE) Conference
  Series

\bibitem[{{Bennett} {et~al.}(2013){Bennett}, {Larson}, {Weiland}, {Jarosik},
  {Hinshaw}, {Odegard}, {Smith}, {Hill}, {Gold}, {Halpern}, {Komatsu}, {Nolta},
  {Page}, {Spergel}, {Wollack}, {Dunkley}, {Kogut}, {Limon}, {Meyer}, {Tucker},
  \& {Wright}}]{wmapfinal}
{Bennett}, C.~L., {Larson}, D., {Weiland}, J.~L., {et~al.} 2013, \apjs, 208, 20

\bibitem[{{Bond} \& {Szalay}(1983)}]{collisionless_damping}
{Bond}, J.~R., \& {Szalay}, A.~S. 1983, \apj, 274, 443

\bibitem[{{Chon} {et~al.}(2004){Chon}, {Challinor}, {Prunet}, {Hivon}, \&
  {Szapudi}}]{polspice}
{Chon}, G., {Challinor}, A., {Prunet}, S., {Hivon}, E., \& {Szapudi}, I. 2004,
  \mnras, 350, 914

\bibitem[{Chuss {et~al.}(2012)Chuss, Wollack, Henry, Hui, Juarez, Krejny,
  Moseley, \& Novak}]{Chuss:12}
Chuss, D.~T., Wollack, E.~J., Henry, R., {et~al.} 2012, Appl. Opt., 51, 197.
\newblock \url{http://ao.osa.org/abstract.cfm?URI=ao-51-2-197}

\bibitem[{{Couchot} {et~al.}(2017){Couchot}, {Henrot-Versill{\'e}},
  {Perdereau}, {Plaszczynski}, {Rouill{\'e} d'Orfeuil}, {Spinelli}, \&
  {Tristram}}]{couchot_neutrino}
{Couchot}, F., {Henrot-Versill{\'e}}, S., {Perdereau}, O., {et~al.} 2017, \aap,
  606, A104

\bibitem[{Dalc{\'{\i}}n {et~al.}(2011)Dalc{\'{\i}}n, Paz, Kler, \&
  Cosimo}]{mpi4py3}
Dalc{\'{\i}}n, L.~D., Paz, R.~R., Kler, P.~A., \& Cosimo, A. 2011, Advances in
  Water Resources, 34, 1124.
\newblock \url{https://doi.org/10.1016/j.advwatres.2011.04.013}

\bibitem[{{DES Collaboration} {et~al.}(2017){DES Collaboration}, {Abbott},
  {Abdalla}, {Alarcon}, {Aleksi{\'c}}, {Allam}, {Allen}, {Amara}, {Annis},
  {Asorey}, {Avila}, {Bacon}, {Balbinot}, {Banerji}, {Banik}, {Barkhouse},
  {Baumer}, {Baxter}, {Bechtol}, {Becker}, {Benoit-L{\'e}vy}, {Benson},
  {Bernstein}, {Bertin}, {Blazek}, {Bridle}, {Brooks}, {Brout}, {Buckley-Geer},
  {Burke}, {Busha}, {Capozzi}, {Carnero Rosell}, {Carrasco Kind}, {Carretero},
  {Castander}, {Cawthon}, {Chang}, {Chen}, {Childress}, {Choi}, {Conselice},
  {Crittenden}, {Crocce}, {Cunha}, {D'Andrea}, {da Costa}, {Das}, {Davis},
  {Davis}, {De Vicente}, {DePoy}, {DeRose}, {Desai}, {Diehl}, {Dietrich},
  {Dodelson}, {Doel}, {Drlica-Wagner}, {Eifler}, {Elliott}, {Elsner},
  {Elvin-Poole}, {Estrada}, {Evrard}, {Fang}, {Fernandez}, {Fert{\'e}},
  {Finley}, {Flaugher}, {Fosalba}, {Friedrich}, {Frieman},
  {Garc{\'{\i}}a-Bellido}, {Garcia-Fernandez}, {Gatti}, {Gaztanaga}, {Gerdes},
  {Giannantonio}, {Gill}, {Glazebrook}, {Goldstein}, {Gruen}, {Gruendl},
  {Gschwend}, {Gutierrez}, {Hamilton}, {Hartley}, {Hinton}, {Honscheid},
  {Hoyle}, {Huterer}, {Jain}, {James}, {Jarvis}, {Jeltema}, {Johnson},
  {Johnson}, {Kacprzak}, {Kent}, {Kim}, {King}, {Kirk}, {Kokron}, {Kovacs},
  {Krause}, {Krawiec}, {Kremin}, {Kuehn}, {Kuhlmann}, {Kuropatkin}, {Lacasa},
  {Lahav}, {Li}, {Liddle}, {Lidman}, {Lima}, {Lin}, {MacCrann}, {Maia},
  {Makler}, {Manera}, {March}, {Marshall}, {Martini}, {McMahon}, {Melchior},
  {Menanteau}, {Miquel}, {Miranda}, {Mudd}, {Muir}, {M{\"o}ller}, {Neilsen},
  {Nichol}, {Nord}, {Nugent}, {Ogando}, {Palmese}, {Peacock}, {Peiris},
  {Peoples}, {Percival}, {Petravick}, {Plazas}, {Porredon}, {Prat}, {Pujol},
  {Rau}, {Refregier}, {Ricker}, {Roe}, {Rollins}, {Romer}, {Roodman},
  {Rosenfeld}, {Ross}, {Rozo}, {Rykoff}, {Sako}, {Salvador}, {Samuroff},
  {S{\'a}nchez}, {Sanchez}, {Santiago}, {Scarpine}, {Schindler}, {Scolnic},
  {Secco}, {Serrano}, {Sevilla-Noarbe}, {Sheldon}, {Smith}, {Smith}, {Smith},
  {Soares-Santos}, {Sobreira}, {Suchyta}, {Tarle}, {Thomas}, {Troxel},
  {Tucker}, {Tucker}, {Uddin}, {Varga}, {Vielzeuf}, {Vikram}, {Vivas},
  {Walker}, {Wang}, {Wechsler}, {Weller}, {Wester}, {Wolf}, {Yanny}, {Yuan},
  {Zenteno}, {Zhang}, {Zhang}, \& {Zuntz}}]{des1yr}
{DES Collaboration}, {Abbott}, T.~M.~C., {Abdalla}, F.~B., {et~al.} 2017, ArXiv
  e-prints, arXiv:1708.01530

\bibitem[{{Eimer} {et~al.}(2012){Eimer}, {Bennett}, {Chuss}, {Marriage},
  {Wollack}, \& {Zeng}}]{class}
{Eimer}, J.~R., {Bennett}, C.~L., {Chuss}, D.~T., {et~al.} 2012, in Society of
  Photo-Optical Instrumentation Engineers (SPIE) Conference Series, Vol. 8452,
  Society of Photo-Optical Instrumentation Engineers (SPIE) Conference Series

\bibitem[{{Einhorn} \& {Sato}(1981)}]{einhorn}
{Einhorn}, M.~B., \& {Sato}, K. 1981, Nuclear Physics B, 180, 385

\bibitem[{{Errard} {et~al.}(2016){Errard}, {Feeney}, {Peiris}, \&
  {Jaffe}}]{cmb4cast}
{Errard}, J., {Feeney}, S.~M., {Peiris}, H.~V., \& {Jaffe}, A.~H. 2016, Journal
  of Cosmology and Astro-Particle Physics, 2016,
  doi:10.1088/1475-7516/2016/03/052

\bibitem[{{Essinger-Hileman} {et~al.}(2014){Essinger-Hileman}, {Ali}, {Amiri},
  {Appel}, {Araujo}, {Bennett}, {Boone}, {Chan}, {Cho}, {Chuss}, {Colazo},
  {Crowe}, {Denis}, {D{\"u}nner}, {Eimer}, {Gothe}, {Halpern}, {Harrington},
  {Hilton}, {Hinshaw}, {Huang}, {Irwin}, {Jones}, {Karakla}, {Kogut}, {Larson},
  {Limon}, {Lowry}, {Marriage}, {Mehrle}, {Miller}, {Miller}, {Moseley},
  {Novak}, {Reintsema}, {Rostem}, {Stevenson}, {Towner}, {U-Yen}, {Wagner},
  {Watts}, {Wollack}, {Xu}, \& {Zeng}}]{class_spie}
{Essinger-Hileman}, T., {Ali}, A., {Amiri}, M., {et~al.} 2014, in Society of
  Photo-Optical Instrumentation Engineers (SPIE) Conference Series, Vol. 9153,
  Society of Photo-Optical Instrumentation Engineers (SPIE) Conference Series

\bibitem[{{Fan} {et~al.}(2006){Fan}, {Strauss}, {Becker}, {White}, {Gunn},
  {Knapp}, {Richards}, {Schneider}, {Brinkmann}, \& {Fukugita}}]{fan}
{Fan}, X., {Strauss}, M.~A., {Becker}, R.~H., {et~al.} 2006, \aj, 132, 117

\bibitem[{{Fendt} \& {Wandelt}(2007)}]{pico}
{Fendt}, W.~A., \& {Wandelt}, B.~D. 2007, \apj, 654, 2

\bibitem[{{Foreman-Mackey}(2016)}]{corner}
{Foreman-Mackey}, D. 2016, The Journal of Open Source Software, 2016,
  doi:10.21105/joss.00024

\bibitem[{{Foreman-Mackey} {et~al.}(2013){Foreman-Mackey}, {Hogg}, {Lang}, \&
  {Goodman}}]{emcee}
{Foreman-Mackey}, D., {Hogg}, D.~W., {Lang}, D., \& {Goodman}, J. 2013, \pasp,
  125, 306

\bibitem[{{Fuskeland} {et~al.}(2014){Fuskeland}, {Wehus}, {Eriksen}, \&
  {N{\ae}ss}}]{specvar}
{Fuskeland}, U., {Wehus}, I.~K., {Eriksen}, H.~K., \& {N{\ae}ss}, S.~K. 2014,
  \apj, 790, 104

\bibitem[{{Galli} {et~al.}(2014){Galli}, {Benabed}, {Bouchet}, {Cardoso},
  {Elsner}, {Hivon}, {Mangilli}, {Prunet}, \& {Wandelt}}]{Galli2014}
{Galli}, S., {Benabed}, K., {Bouchet}, F., {et~al.} 2014, \prd, 90, 063504

\bibitem[{{Gandilo} {et~al.}(2016){Gandilo}, {Ade}, {Benford}, {Bennett},
  {Chuss}, {Dotson}, {Eimer}, {Fixsen}, {Halpern}, {Hilton}, {Hinshaw},
  {Irwin}, {Jhabvala}, {Kimball}, {Kogut}, {Lowe}, {McMahon}, {Miller},
  {Mirel}, {Moseley}, {Pawlyk}, {Rodriguez}, {Sharp}, {Shirron}, {Staguhn},
  {Sullivan}, {Switzer}, {Taraschi}, {Tucker}, \& {Wollack}}]{gandilo}
{Gandilo}, N.~N., {Ade}, P.~A.~R., {Benford}, D., {et~al.} 2016, in \procspie,
  Vol. 9914, Millimeter, Submillimeter, and Far-Infrared Detectors and
  Instrumentation for Astronomy VIII, 99141J

\bibitem[{{G{\'o}rski} {et~al.}(2005){G{\'o}rski}, {Hivon}, {Banday},
  {Wandelt}, {Hansen}, {Reinecke}, \& {Bartelmann}}]{healpix}
{G{\'o}rski}, K.~M., {Hivon}, E., {Banday}, A.~J., {et~al.} 2005, \apj, 622,
  759

\bibitem[{{Gunn} \& {Peterson}(1965)}]{gunnpeterson}
{Gunn}, J.~E., \& {Peterson}, B.~A. 1965, \apj, 142, 1633

\bibitem[{{Guth}(1981)}]{guth}
{Guth}, A.~H. 1981, \prd, 23, 347

\bibitem[{{Hamimeche} \& {Lewis}(2008)}]{wishart}
{Hamimeche}, S., \& {Lewis}, A. 2008, \prd, 77, 103013

\bibitem[{{Harrington} {et~al.}(2016){Harrington}, {Marriage}, {Ali}, {Appel},
  {Bennett}, {Boone}, {Brewer}, {Chan}, {Chuss}, {Colazo}, {Dahal}, {Denis},
  {D{\"u}nner}, {Eimer}, {Essinger-Hileman}, {Fluxa}, {Halpern}, {Hilton},
  {Hinshaw}, {Hubmayr}, {Iuliano}, {Karakla}, {McMahon}, {Miller}, {Moseley},
  {Palma}, {Parker}, {Petroff}, {Pradenas}, {Rostem}, {Sagliocca}, {Valle},
  {Watts}, {Wollack}, {Xu}, \& {Zeng}}]{2016SPIE.9914E..1KH}
{Harrington}, K., {Marriage}, T., {Ali}, A., {et~al.} 2016, in \procspie, Vol.
  9914, Millimeter, Submillimeter, and Far-Infrared Detectors and
  Instrumentation for Astronomy VIII, 99141K

\bibitem[{{Heinrich} {et~al.}(2017){Heinrich}, {Miranda}, \&
  {Hu}}]{2017PhRvD..95b3513H}
{Heinrich}, C.~H., {Miranda}, V., \& {Hu}, W. 2017, \prd, 95, 023513

\bibitem[{{Henning} {et~al.}(2017){Henning}, {Sayre}, {Reichardt}, {Ade},
  {Anderson}, {Austermann}, {Beall}, {Bender}, {Benson}, {Bleem}, {Carlstrom},
  {Chang}, {Chiang}, {Cho}, {Citron}, {Corbett Moran}, {Crawford}, {Crites},
  {de Haan}, {Dobbs}, {Everett}, {Gallicchio}, {George}, {Gilbert},
  {Halverson}, {Harrington}, {Hilton}, {Holder}, {Holzapfel}, {Hoover}, {Hou},
  {Hrubes}, {Huang}, {Hubmayr}, {Irwin}, {Keisler}, {Knox}, {Lee}, {Leitch},
  {Li}, {Lowitz}, {Manzotti}, {McMahon}, {Meyer}, {Mocanu}, {Montgomery},
  {Nadolski}, {Natoli}, {Nibarger}, {Novosad}, {Padin}, {Pryke}, {Ruhl},
  {Saliwanchik}, {Schaffer}, {Sievers}, {Smecher}, {Stark}, {Story}, {Tucker},
  {Vanderlinde}, {Veach}, {Vieira}, {Wang}, {Whitehorn}, {Wu}, \&
  {Yefremenko}}]{sptEmode}
{Henning}, J.~W., {Sayre}, J.~T., {Reichardt}, C.~L., {et~al.} 2017, ArXiv
  e-prints, arXiv:1707.09353

\bibitem[{{Hu} \& {Okamoto}(2004)}]{principalpower}
{Hu}, W., \& {Okamoto}, T. 2004, \prd, 69, 043004

\bibitem[{{Hu} \& {Sugiyama}(1996)}]{analytic_damping}
{Hu}, W., \& {Sugiyama}, N. 1996, \apj, 471, 542

\bibitem[{Hunter(2007)}]{matplotlib}
Hunter, J.~D. 2007, Computing In Science \& Engineering, 9, 90

\bibitem[{Jones {et~al.}(2001--)Jones, Oliphant, Peterson, {et~al.}}]{scipy}
Jones, E., Oliphant, T., Peterson, P., {et~al.} 2001--, {SciPy}: Open source
  scientific tools for {Python}, , .
\newblock \url{http://www.scipy.org/}

\bibitem[{{Kamionkowski} {et~al.}(1997){Kamionkowski}, {Kosowsky}, \&
  {Stebbins}}]{kks}
{Kamionkowski}, M., {Kosowsky}, A., \& {Stebbins}, A. 1997, \prd, 55, 7368

\bibitem[{{Kazanas}(1980)}]{kazanas}
{Kazanas}, D. 1980, \apjl, 241, L59

\bibitem[{{Krachmalnicoff} {et~al.}(2016){Krachmalnicoff}, {Baccigalupi},
  {Aumont}, {Bersanelli}, \& {Mennella}}]{2016A&A...588A..65K}
{Krachmalnicoff}, N., {Baccigalupi}, C., {Aumont}, J., {Bersanelli}, M., \&
  {Mennella}, A. 2016, \aap, 588, A65

\bibitem[{{Lewis} {et~al.}(2000){Lewis}, {Challinor}, \& {Lasenby}}]{camb}
{Lewis}, A., {Challinor}, A., \& {Lasenby}, A. 2000, \apj, 538, 473

\bibitem[{{Linde}(1982)}]{linde}
{Linde}, A.~D. 1982, Physics Letters B, 108, 389

\bibitem[{{Liu} {et~al.}(2016){Liu}, {Pritchard}, {Allison}, {Parsons},
  {Seljak}, \& {Sherwin}}]{Liu2015}
{Liu}, A., {Pritchard}, J.~R., {Allison}, R., {et~al.} 2016, \prd, 93, 043013

\bibitem[{{Louis} {et~al.}(2017){Louis}, {Grace}, {Hasselfield}, {Lungu},
  {Maurin}, {Addison}, {Ade}, {Aiola}, {Allison}, {Amiri}, {Angile},
  {Battaglia}, {Beall}, {de Bernardis}, {Bond}, {Britton}, {Calabrese}, {Cho},
  {Choi}, {Coughlin}, {Crichton}, {Crowley}, {Datta}, {Devlin}, {Dicker},
  {Dunkley}, {D{\"u}nner}, {Ferraro}, {Fox}, {Gallardo}, {Gralla}, {Halpern},
  {Henderson}, {Hill}, {Hilton}, {Hilton}, {Hincks}, {Hlozek}, {Ho}, {Huang},
  {Hubmayr}, {Huffenberger}, {Hughes}, {Infante}, {Irwin}, {Muya Kasanda},
  {Klein}, {Koopman}, {Kosowsky}, {Li}, {Madhavacheril}, {Marriage}, {McMahon},
  {Menanteau}, {Moodley}, {Munson}, {Naess}, {Nati}, {Newburgh}, {Nibarger},
  {Niemack}, {Nolta}, {Nu{\~n}ez}, {Page}, {Pappas}, {Partridge}, {Rojas},
  {Schaan}, {Schmitt}, {Sehgal}, {Sherwin}, {Sievers}, {Simon}, {Spergel},
  {Staggs}, {Switzer}, {Thornton}, {Trac}, {Treu}, {Tucker}, {Van Engelen},
  {Ward}, \& {Wollack}}]{actpollens}
{Louis}, T., {Grace}, E., {Hasselfield}, M., {et~al.} 2017, \jcap, 6, 031

\bibitem[{{Mangilli} {et~al.}(2015){Mangilli}, {Plaszczynski}, \&
  {Tristram}}]{cross_cl}
{Mangilli}, A., {Plaszczynski}, S., \& {Tristram}, M. 2015, \mnras, 453, 3174

\bibitem[{{Miller} {et~al.}(2016){Miller}, {Chuss}, {Marriage}, {Wollack},
  {Appel}, {Bennett}, {Eimer}, {Essinger-Hileman}, {Fixsen}, {Harrington},
  {Moseley}, {Rostem}, {Switzer}, \& {Watts}}]{miller}
{Miller}, N.~J., {Chuss}, D.~T., {Marriage}, T.~A., {et~al.} 2016, \apj, 818,
  151

\bibitem[{{Mukhanov} \& {Chibisov}(1981)}]{mukhanov}
{Mukhanov}, V.~F., \& {Chibisov}, G.~V. 1981, Soviet Journal of Experimental
  and Theoretical Physics Letters, 33, 532

\bibitem[{{Page} {et~al.}(2007){Page}, {Hinshaw}, {Komatsu}, {Nolta},
  {Spergel}, {Bennett}, {Barnes}, {Bean}, {Dor{\'e}}, {Dunkley}, {Halpern},
  {Hill}, {Jarosik}, {Kogut}, {Limon}, {Meyer}, {Odegard}, {Peiris}, {Tucker},
  {Verde}, {Weiland}, {Wollack}, \& {Wright}}]{pol_mask}
{Page}, L., {Hinshaw}, G., {Komatsu}, E., {et~al.} 2007, \apjs, 170, 335

\bibitem[{Patrignani {et~al.}(2016)}]{pdg17}
Patrignani, C., {et~al.} 2016, Chin. Phys., C40, 100001

\bibitem[{P\'erez \& Granger(2007)}]{ipython}
P\'erez, F., \& Granger, B.~E. 2007, Computing in Science and Engineering, 9,
  21.
\newblock \url{http://ipython.org}

\bibitem[{{Planck Collaboration} {et~al.}(2016){Planck Collaboration},
  {Aghanim}, {Ashdown}, {Aumont}, {Baccigalupi}, {Ballardini}, {Banday},
  {Barreiro}, {Bartolo}, {Basak}, {Battye}, {Benabed}, {Bernard}, {Bersanelli},
  {Bielewicz}, {Bock}, {Bonaldi}, {Bonavera}, {Bond}, {Borrill}, {Bouchet},
  {Boulanger}, {Bucher}, {Burigana}, {Butler}, {Calabrese}, {Cardoso},
  {Carron}, {Challinor}, {Chiang}, {Colombo}, {Combet}, {Comis}, {Coulais},
  {Crill}, {Curto}, {Cuttaia}, {Davis}, {de Bernardis}, {de Rosa}, {de Zotti},
  {Delabrouille}, {Delouis}, {Di Valentino}, {Dickinson}, {Diego}, {Dor{\'e}},
  {Douspis}, {Ducout}, {Dupac}, {Efstathiou}, {Elsner}, {En{\ss}lin},
  {Eriksen}, {Falgarone}, {Fantaye}, {Finelli}, {Forastieri}, {Frailis},
  {Fraisse}, {Franceschi}, {Frolov}, {Galeotta}, {Galli}, {Ganga},
  {G{\'e}nova-Santos}, {Gerbino}, {Ghosh}, {Gonz{\'a}lez-Nuevo}, {G{\'o}rski},
  {Gratton}, {Gruppuso}, {Gudmundsson}, {Hansen}, {Helou},
  {Henrot-Versill{\'e}}, {Herranz}, {Hivon}, {Huang}, {Ili{\'c}}, {Jaffe},
  {Jones}, {Keih{\"a}nen}, {Keskitalo}, {Kisner}, {Knox}, {Krachmalnicoff},
  {Kunz}, {Kurki-Suonio}, {Lagache}, {Lamarre}, {Langer}, {Lasenby},
  {Lattanzi}, {Lawrence}, {Le Jeune}, {Leahy}, {Levrier}, {Liguori}, {Lilje},
  {L{\'o}pez-Caniego}, {Ma}, {Mac{\'{\i}}as-P{\'e}rez}, {Maggio}, {Mangilli},
  {Maris}, {Martin}, {Mart{\'{\i}}nez-Gonz{\'a}lez}, {Matarrese}, {Mauri},
  {McEwen}, {Meinhold}, {Melchiorri}, {Mennella}, {Migliaccio},
  {Miville-Desch{\^e}nes}, {Molinari}, {Moneti}, {Montier}, {Morgante}, {Moss},
  {Mottet}, {Naselsky}, {Natoli}, {Oxborrow}, {Pagano}, {Paoletti},
  {Partridge}, {Patanchon}, {Patrizii}, {Perdereau}, {Perotto}, {Pettorino},
  {Piacentini}, {Plaszczynski}, {Polastri}, {Polenta}, {Puget}, {Rachen},
  {Racine}, {Reinecke}, {Remazeilles}, {Renzi}, {Rocha}, {Rossetti}, {Roudier},
  {Rubi{\~n}o-Mart{\'{\i}}n}, {Ruiz-Granados}, {Salvati}, {Sandri},
  {Savelainen}, {Scott}, {Sirri}, {Sunyaev}, {Suur-Uski}, {Tauber}, {Tenti},
  {Toffolatti}, {Tomasi}, {Tristram}, {Trombetti}, {Valiviita}, {Van Tent},
  {Vibert}, {Vielva}, {Villa}, {Vittorio}, {Wandelt}, {Watson}, {Wehus},
  {White}, {Zacchei}, \& {Zonca}}]{planck_redsys}
{Planck Collaboration}, {Aghanim}, N., {Ashdown}, M., {et~al.} 2016, \aap, 596,
  A107

\bibitem[{{Planck Collaboration} {et~al.}(2018){Planck Collaboration},
  {Akrami}, {Ashdown}, {Aumont}, {Baccigalupi}, {Ballardini}, {Banday},
  {Barreiro}, {Bartolo}, {Basak}, {Benabed}, {Bernard}, {Bersanelli},
  {Bielewicz}, {Bond}, {Borrill}, {Bouchet}, {Boulanger}, {Bracco}, {Bucher},
  {Burigana}, {Calabrese}, {Cardoso}, {Carron}, {Chiang}, {Combet}, {Crill},
  {de Bernardis}, {de Zotti}, {Delabrouille}, {Delouis}, {Di Valentino},
  {Dickinson}, {Diego}, {Ducout}, {Dupac}, {Elsner}, {En{\ss}lin}, {Falgarone},
  {Fantaye}, {Ferri{\`e}re}, {Finelli}, {Forastieri}, {Frailis}, {Fraisse},
  {Franceschi}, {Frolov}, {Galeotta}, {Galli}, {Ganga}, {G{\'e}nova-Santos},
  {Ghosh}, {Gonz{\'a}lez-Nuevo}, {G{\'o}rski}, {Gruppuso}, {Gudmundsson},
  {Guillet}, {Handley}, {Hansen}, {Herranz}, {Huang}, {Jaffe}, {Jones},
  {Keih{\"a}nen}, {Keskitalo}, {Kiiveri}, {Kim}, {Krachmalnicoff}, {Kunz},
  {Kurki-Suonio}, {Lamarre}, {Lasenby}, {Le Jeune}, {Levrier}, {Liguori},
  {Lilje}, {Lindholm}, {L{\'o}pez-Caniego}, {Lubin}, {Ma},
  {Mac{\'{\i}}as-P{\'e}rez}, {Maggio}, {Maino}, {Mandolesi}, {Mangilli},
  {Martin}, {Mart{\'{\i}}nez-Gonz{\'a}lez}, {Matarrese}, {McEwen}, {Meinhold},
  {Melchiorri}, {Migliaccio}, {Miville-Desch{\^e}nes}, {Molinari}, {Moneti},
  {Montier}, {Morgante}, {Natoli}, {Pagano}, {Paoletti}, {Pettorino},
  {Piacentini}, {Polenta}, {Rachen}, {Reinecke}, {Remazeilles}, {Renzi},
  {Rocha}, {Rosset}, {Roudier}, {Rubi{\~n}o-Mart{\'{\i}}n}, {Ruiz-Granados},
  {Salvati}, {Sandri}, {Savelainen}, {Scott}, {Soler}, {Spencer}, {Tauber},
  {Tavagnacco}, {Toffolatti}, {Tomasi}, {Trombetti}, {Valiviita}, {Vansyngel},
  {Van Tent}, {Vielva}, {Villa}, {Vittorio}, {Wehus}, {Zacchei}, \&
  {Zonca}}]{newplanckdust}
{Planck Collaboration}, {Akrami}, Y., {Ashdown}, M., {et~al.} 2018, ArXiv
  e-prints, arXiv:1801.04945

\bibitem[{{Planck Collaboration Int.~L}(2017)}]{planckL2016}
{Planck Collaboration Int.~L}. 2017, \aap, 599, A51

\bibitem[{{Planck Collaboration Int.~XLVII}(2016)}]{plancktau}
{Planck Collaboration Int.~XLVII}. 2016, \aap, 596, A108

\bibitem[{{Planck Collaboration X}(2016)}]{planckfg}
{Planck Collaboration X}. 2016, \aap, 594, A10

\bibitem[{{Planck Collaboration XI}(2016)}]{plancklike15}
{Planck Collaboration XI}. 2016, \aap, 594, 104

\bibitem[{{Planck Collaboration XIII}(2016)}]{planck}
{Planck Collaboration XIII}. 2016, \aap, 594, A13

\bibitem[{{Seljak} \& {Zaldarriaga}(1997)}]{seljzald}
{Seljak}, U., \& {Zaldarriaga}, M. 1997, Physical Review Letters, 78, 2054

\bibitem[{{Sheehy} \& {Slosar}(2017)}]{sheehy}
{Sheehy}, C., \& {Slosar}, A. 2017, ArXiv e-prints, arXiv:1709.09729

\bibitem[{{Smith} \& {Zaldarriaga}(2007)}]{purepseudocl}
{Smith}, K.~M., \& {Zaldarriaga}, M. 2007, \prd, 76, 043001

\bibitem[{{Starobinsky}(1980)}]{starobinsky}
{Starobinsky}, A.~A. 1980, Physics Letters B, 91, 99

\bibitem[{{Suzuki} {et~al.}(2016){Suzuki}, {Ade}, {Akiba}, {Aleman}, {Arnold},
  {Baccigalupi}, {Barch}, {Barron}, {Bender}, {Boettger}, {Borrill}, {Chapman},
  {Chinone}, {Cukierman}, {Dobbs}, {Ducout}, {Dunner}, {Elleflot}, {Errard},
  {Fabbian}, {Feeney}, {Feng}, {Fujino}, {Fuller}, {Gilbert}, {Goeckner-Wald},
  {Groh}, {Haan}, {Hall}, {Halverson}, {Hamada}, {Hasegawa}, {Hattori},
  {Hazumi}, {Hill}, {Holzapfel}, {Hori}, {Howe}, {Inoue}, {Irie}, {Jaehnig},
  {Jaffe}, {Jeong}, {Katayama}, {Kaufman}, {Kazemzadeh}, {Keating}, {Kermish},
  {Keskitalo}, {Kisner}, {Kusaka}, {Jeune}, {Lee}, {Leon}, {Linder}, {Lowry},
  {Matsuda}, {Matsumura}, {Miller}, {Mizukami}, {Montgomery}, {Navaroli},
  {Nishino}, {Peloton}, {Poletti}, {Puglisi}, {Rebeiz}, {Raum}, {Reichardt},
  {Richards}, {Ross}, {Rotermund}, {Segawa}, {Sherwin}, {Shirley},
  {Siritanasak}, {Stebor}, {Stompor}, {Suzuki}, {Tajima}, {Takada}, {Takakura},
  {Takatori}, {Tikhomirov}, {Tomaru}, {Westbrook}, {Whitehorn}, {Yamashita},
  {Zahn}, \& {Zahn}}]{polarbear}
{Suzuki}, A., {Ade}, P., {Akiba}, Y., {et~al.} 2016, Journal of Low Temperature
  Physics, 184, 805

\bibitem[{{Tajima} {et~al.}(2012){Tajima}, {Choi}, {Hazumi}, {Ishitsuka},
  {Kawai}, \& {Yoshida}}]{groundbird}
{Tajima}, O., {Choi}, J., {Hazumi}, M., {et~al.} 2012, in Society of
  Photo-Optical Instrumentation Engineers (SPIE) Conference Series, Vol. 8452,
  Society of Photo-Optical Instrumentation Engineers (SPIE) Conference Series

\bibitem[{{Thorne} {et~al.}(2017){Thorne}, {Dunkley}, {Alonso}, \&
  {N{\ae}ss}}]{pysm}
{Thorne}, B., {Dunkley}, J., {Alonso}, D., \& {N{\ae}ss}, S. 2017, \mnras, 469,
  2821

\bibitem[{{Thornton} {et~al.}(2016){Thornton}, {Ade}, {Aiola}, {Angil{\`e}},
  {Amiri}, {Beall}, {Becker}, {Cho}, {Choi}, {Corlies}, {Coughlin}, {Datta},
  {Devlin}, {Dicker}, {D{\"u}nner}, {Fowler}, {Fox}, {Gallardo}, {Gao},
  {Grace}, {Halpern}, {Hasselfield}, {Henderson}, {Hilton}, {Hincks}, {Ho},
  {Hubmayr}, {Irwin}, {Klein}, {Koopman}, {Li}, {Louis}, {Lungu}, {Maurin},
  {McMahon}, {Munson}, {Naess}, {Nati}, {Newburgh}, {Nibarger}, {Niemack},
  {Niraula}, {Nolta}, {Page}, {Pappas}, {Schillaci}, {Schmitt}, {Sehgal},
  {Sievers}, {Simon}, {Staggs}, {Tucker}, {Uehara}, {van Lanen}, {Ward}, \&
  {Wollack}}]{actpol}
{Thornton}, R.~J., {Ade}, P.~A.~R., {Aiola}, S., {et~al.} 2016, \apjs, 227, 21

\bibitem[{van~der Walt {et~al.}(2011)van~der Walt, Colbert, \&
  Varoquaux}]{numpy}
van~der Walt, S., Colbert, S.~C., \& Varoquaux, G. 2011, Computing in Science
  {\&} Engineering, 13, 22.
\newblock \url{https://doi.org/10.1109%2Fmcse.2011.37}

\bibitem[{Watts(2017)}]{cleefast}
Watts, D.~J. 2017, \texttt{clee\_fast},  Zenodo, doi:10.5281/zenodo.1066547.
\newblock \url{http://dx.doi.org/10.5281/zenodo.1066547}

\bibitem[{{Watts} {et~al.}(2015){Watts}, {Larson}, {Marriage}, {Abitbol},
  {Appel}, {Bennett}, {Chuss}, {Eimer}, {Essinger-Hileman}, {Miller}, {Rostem},
  \& {Wollack}}]{lowellpix}
{Watts}, D.~J., {Larson}, D., {Marriage}, T.~A., {et~al.} 2015, \apj, 814, 103

\bibitem[{{Weiland} {et~al.}(2018){Weiland}, {Osumi}, {Addison}, {Bennett},
  {Watts}, {Halpern}, \& {Hinshaw}}]{weiland}
{Weiland}, J.~L., {Osumi}, K., {Addison}, G.~E., {et~al.} 2018, ArXiv e-prints,
  arXiv:1801.01226

\bibitem[{{Wu} {et~al.}(2016){Wu}, {Ade}, {Ahmed}, {Alexander}, {Amiri},
  {Barkats}, {Benton}, {Bischoff}, {Bock}, {Bowens-Rubin}, {Buder}, {Bullock},
  {Buza}, {Connors}, {Filippini}, {Fliescher}, {Grayson}, {Halpern},
  {Harrison}, {Hilton}, {Hristov}, {Hui}, {Irwin}, {Kang}, {Karkare}, {Karpel},
  {Kefeli}, {Kernasovskiy}, {Kovac}, {Kuo}, {Megerian}, {Netterfield},
  {Nguyen}, {O'Brient}, {Ogburn}, {Pryke}, {Reintsema}, {Richter}, {Sorensen},
  {Staniszewski}, {Steinbach}, {Sudiwala}, {Teply}, {Thompson}, {Tolan},
  {Tucker}, {Turner}, {Vieregg}, {Weber}, {Wiebe}, {Willmert}, \&
  {Yoon}}]{bicep/keck}
{Wu}, W.~L.~K., {Ade}, P.~A.~R., {Ahmed}, Z., {et~al.} 2016, Journal of Low
  Temperature Physics, 184, 765

\end{thebibliography}
